\documentclass[%
reprint,
superscriptaddress,
bibnotes,
amsmath,amssymb,
prl,
floatfix,
]{revtex4-2}
\usepackage{graphicx}
\usepackage{dcolumn}
\usepackage{bm}
\usepackage[breaklinks=true,hypertexnames=false]{hyperref}
\usepackage{natbib}
\usepackage{mathrsfs,mathtools,wasysym,dsfont,here}
\usepackage{amsthm}
\usepackage{physics}
\usepackage{amsfonts}
\usepackage{array,booktabs}
\usepackage{comment}
\usepackage{cases}
\usepackage{bbold}

\usepackage[usenames,dvipsnames]{xcolor}
\usepackage{tikz}
\usepackage{pgfplots}
\usetikzlibrary{shapes.geometric, arrows.meta, positioning, calc, decorations.markings, decorations.pathreplacing,calligraphy}
\pgfplotsset{compat=1.18}

\usetikzlibrary{calc,decorations.pathreplacing,decorations.pathmorphing}

\tikzset{
	 edge/.style={thick, draw=Gray},
}

\newcommand{\ad}{\operatorname{ad}}

\newcommand{\Ker}{\operatorname{Ker}}

\theoremstyle{plain}

\theoremstyle{definition}

\theoremstyle{remark}

\def\linkcolor{ForestGreen}

\hypersetup{
    colorlinks=true, urlcolor=\linkcolor,citecolor=\linkcolor,linkcolor=\linkcolor
}

\newcommand{\secc}[1]{\textit{#1}.---}

\newcommand{\Li}{{\mathcal{L}}}
\newcommand{\Hi}{{\mathcal{H}}}
\newcommand{\imi}{\mathrm{i}}

\newcommand{\jbm}{\boldsymbol{j}}

\newcommand{\ellbm}{\boldsymbol{\ell}}

\newcommand{\poly}[2]{P_{#1}(#2)}

\newcommand{\polytilpm}[3]{\widetilde{P}_{#1}^{#2}(#3)}

\newcommand{\indepset}[1]{\mathcal{S}_{#1}}

\newcommand{\TODO}[1]{\ifmmode\text{\color{red}\textbf{TODO:} #1}\else\textcolor{red}{\textbf{TODO:} #1}\fi}

\newcommand{\liouvillian}[1]{\widetilde{#1}}

\newcommand{\Psitil}{\liouvillian{\Psi}}
\newcommand{\Ntil}{\liouvillian{\mathcal N}}
\newcommand{\md}{\text{-}}

\newcommand{\tokyo}{\affiliation{Department of Physics, Graduate School of Science, The University of Tokyo, \\ 7-3-1, Hongo, Bunkyo-ku, Tokyo, 113-0033, Japan}}
\newcommand{\tokyoIntelligence}{\affiliation{Institute for Physics of Intelligence, The University of Tokyo, 7-3-1 Hongo, Tokyo 113-0033, Japan}}
\newcommand{\tokyoTrans}{\affiliation{Trans-Scale Quantum Science Institute, The University of Tokyo, 7-3-1, Hongo, Tokyo 113-0033, Japan}}
\newcommand{\riken}{\affiliation{Nonequilibrium Quantum Statistical Mechanics RIKEN Hakubi Research Team,
RIKEN Pioneering Research Institute (PRI), Wako, Saitama 351-0198, Japan}}

\preprint{APS/123-QED}

\begin{document}

\title{Dissipative free fermions in disguise}

\author{Kohei Fukai}
\tokyo
\email{kohei.fukai@phys.s.u-tokyo.ac.jp}
\author{Hironobu Yoshida}
\tokyo
\riken
\author{Hosho Katsura}
\tokyo
\tokyoIntelligence
\tokyoTrans

\begin{abstract}
    \noindent
    Recently, a class of spin chains known as ``free fermions in disguise'' (FFD) has been discovered, which possess hidden free-fermion spectra even though they are not solvable via the standard Jordan-Wigner transformation.
    In this work, we extend this FFD framework to open quantum systems governed by the Gorini-Kossakowski-Sudarshan-Lindblad (GKSL) equation.
    We establish a general class of exactly solvable open quantum systems within the FFD framework: if the Liouvillian frustration graph is claw-free and has a simplicial clique, the Liouvillian possesses a hidden free-fermion spectrum.
    In particular, the \textit{(even-hole, claw)-free} condition automatically guarantees this, enabling exact computation of the Liouvillian gap and an infinite-temperature autocorrelation function.
    Our results provide the first realization of the FFD mechanism in open quantum systems.
\end{abstract}

\maketitle

\secc{Introduction}
Exact solutions play a pivotal role in understanding the dynamics of quantum many-body systems.
While the generalized Gibbs ensemble and generalized hydrodynamics describe the equilibration and transport in integrable models \cite{rigol-gge,JS-CGGE,doyon-ghd,jacopo-ghd}, constructing models where the full spectrum and dynamics can be accessed exactly remains a central challenge.

Recently, Fendley constructed a spin chain that is quartic in Jordan-Wigner fermions yet possesses a hidden free-fermion spectrum~\cite{FFD-fendley-2019}, now known as ``free fermions in disguise'' (FFD) \cite{alcaraz-pimenta-multispin-2020,alcaraz-pimenta-spectrum-2020,FFD-chapman-2021,FFD-chapman-unified-2023,alcaraz-pimenta-sirker-2023,fendley-pozsgay-2024,Istvan-ffd-prb-2025,mann-elman-wood-chapman-parafermion-2025,fukai-pozsgay-ffd-circuit-2025,fukai-claw-2025,alcaraz-nonhomogeneous-range-2025,Eric-FFD-hilbert-space-2025,David-ffd-circ-2025,fukai-vona-pozsgay-ffd-eigenstates-paper1-2026,Ruh-Elman-2026-twin-collapse}.
Unlike conventional free-fermion-solvable spin models~\cite{LSM,pfeuty1970,kitaev2006anyons}, FFD models cannot be diagonalized by the Jordan-Wigner transformation \cite{FFD-chapman-2021}.
Free-fermion solvability is now understood in graph-theoretic terms~\cite{ogura-free-fermion-graph-2020,chapman2020characterization,FFD-chapman-2021,FFD-chapman-unified-2023}: the \emph{(even-hole, claw)-free} (ECF) condition on the frustration graph guarantees FFD solvability~\cite{FFD-chapman-2021}, later extended to claw-free graphs with a simplicial clique~\cite{FFD-chapman-unified-2023}.

While these developments have deepened our understanding of solvable closed systems, realistic quantum systems are inevitably coupled to an environment, leading to dissipative dynamics.
When the system-environment coupling is Markovian, the time evolution of the density matrix is governed by the Gorini-Kossakowski-Sudarshan-Lindblad (GKSL) equation \cite{lindblad-eredeti,rivas2010markovian,lindblad-intro}.
Although dissipation generally breaks integrability \cite{Lindblad-noise,vasseur-breaking}, finding exactly solvable Lindbladians is of great importance, as they can serve as benchmarks for approximate methods \cite{zala-tGGE} and offer insights into non-equilibrium steady states (NESS) and relaxation dynamics \cite{enej-thesis,haga-nakagawa-hamazaki-ueda-incoherentons-2023}.

Several strategies have been developed to find exact solutions for the GKSL equation \cite{rowlands-lamacraft-noisy-spins,essler-piroli-lindblad}.
A prominent class consists of quadratic open systems, which can be diagonalized using third quantization techniques \cite{third-quantization,third-quantization-2,katsura-lindblad-1,eric-lindblad}.
In the context of transport, boundary-driven spin chains have been extensively studied, often allowing for the explicit construction of the NESS \cite{prosen-boundary-lindblad-1,prosen-boundary-lindblad-2,boundary-lindblad-mps,prosen-exterior-lindblad,enej-thesis,matsui-prosen-boundary-ness-2017,Popkov-helix-2017,matsui-tsuji-impurity-ness-2024}.
More recently, solvability has been established for superoperators exhibiting a triangular structure \cite{marko-lindblad-1,japanok-lindblad} and for some Yang-Baxter integrable models \cite{essler-prosen-lindblad,katsura-lindblad-2,essler-lindblad-review,Leeuw-PRL-2021,PhysRevResearch.6.L032067}.
Despite this progress, the FFD framework has not yet been extended to open quantum systems.
It remains an open question whether dissipative couplings can be designed such that the Liouvillian superoperator exhibits a hidden free-fermion spectrum.

In this paper, we answer this question affirmatively.
We show that if the Hamiltonian and jump operators are designed so that the Liouvillian frustration graph is claw-free and has a simplicial clique, the Liouvillian is solvable by hidden free fermions.
In particular, we show that any ECF frustration graph yields an exactly solvable dissipative spin chain.
Crucially, this solvability holds for arbitrary coupling constants without fine-tuning, and a non-local mapping beyond the standard Jordan-Wigner transformation is utilized.
We can explicitly derive the Liouvillian spectrum, the finite-size scaling of the Liouvillian gap, and an infinite-temperature autocorrelation function, demonstrating that the FFD mechanism can be successfully extended to the realm of open quantum systems.

\secc{The GKSL master equation}
We consider a spin-$1/2$ system whose Markovian dynamics is described by the Gorini-Kossakowski-Sudarshan-Lindblad (GKSL) master equation \cite{lindblad-eredeti,lindblad-intro}.
The time evolution of the density matrix $\rho(t)$ is given by
\begin{equation}
    \label{eq:lindblad}
    \dot \rho \equiv \Li\rho = -\imi [H, \rho] + \sum_a \Big(
    \ell_a\rho \ell_a^\dagger-\frac{1}{2}\{ \ell^\dagger_a \ell_a,\rho\}\Big).
\end{equation}
Here, $\Li$ is the Liouvillian superoperator, $H$ is the Hamiltonian for the unitary dynamics, and $\ell_a$ are the jump operators describing dissipative processes.

To analyze the spectrum of the Liouvillian, it is convenient to employ the vectorization formalism.
We map the density matrix $\rho$ to a state vector $|\rho\rangle\rangle$ in the doubled Hilbert space $\mathcal{H}\otimes \mathcal{H}$, treating the Liouvillian as a non-Hermitian Hamiltonian acting on the ket and bra spaces following the argument in Refs.~\cite{znidaric-pre-2014,znidaric-pre-2015,katsura-lindblad-1,katsura-lindblad-2,Leeuw-PRL-2021}.
Specifically, operators acting from the left (right) on $\rho$ are mapped to operators acting on the ket (bra) space.
The Liouvillian $\Li$ in the doubled Hilbert space reads
\begin{align}
    \label{eq:liouvillian-doubled}
    \Hi \equiv & \imi \Li = H^{(1)} - H^{(2)*}
    \nonumber                                  \\
               & \hspace{0em}
    + \imi \sum_{a} \qty[ \ell^{(1)}_{a} \ell^{(2)*}_{a} -\frac{1}{2} \ell^{(1)\dagger}_{a} \ell^{(1)}_{a} -\frac{1}{2} \ell^{(2)T}_{a} \ell^{(2)*}_{a} ]
    \,.
\end{align}
Here, $^T$ denotes transpose, $^*$ complex conjugation, and $A^{(1)} = A \otimes I$ (resp.\ $A^{(2)} = I \otimes A$) acts on the ket (resp.\ bra) space.
In this paper, we construct a class of Liouvillians $\Li$ exactly solvable via the FFD framework.

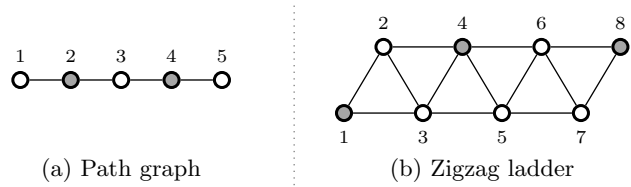
\begin{figure}[tb]
    \centering
    \begin{tikzpicture}[
            vs/.style={circle, draw=black, minimum size=2mm, inner sep=0.5mm, line width=0.4mm},
            vssel/.style={circle, draw=black, fill=black!35, minimum size=2mm, inner sep=0.5mm, line width=0.4mm},
            eg/.style={draw=black, line width=0.5pt}
        ]
        \begin{scope}[shift={(-2.8,0)}]
            \node[vs, label={[font=\scriptsize]above:$1$}] (p1) at ({0.67*(-2)},0) {};
            \node[vssel, label={[font=\scriptsize]above:$2$}] (p2) at ({0.67*(-1)},0) {};
            \node[vs, label={[font=\scriptsize]above:$3$}] (p3) at (0,0) {};
            \node[vssel, label={[font=\scriptsize]above:$4$}] (p4) at ({0.67*1},0) {};
            \node[vs, label={[font=\scriptsize]above:$5$}] (p5) at ({0.67*2},0) {};
            \foreach \i/\j in {1/2,2/3,3/4,4/5} {
                    \draw[eg] (p\i) -- (p\j);
                }
            \node at (0,-1.2) {(a) Path graph};
        \end{scope}
        \draw[dotted, gray, line width=0.5pt] (-0.5,-1.4) -- (-0.5,1.0);
        \begin{scope}[shift={(2.0,0)}]
            \pgfmathsetmacro{\xs}{0.525}
            \pgfmathsetmacro{\ys}{0.44}
            \node[vssel, label={[font=\scriptsize]below:$1$}] (w1) at ({-3.5*\xs},-\ys) {};
            \node[vs, label={[font=\scriptsize]below:$3$}] (w3) at ({-1.5*\xs},-\ys) {};
            \node[vs, label={[font=\scriptsize]below:$5$}] (w5) at ({0.5*\xs},-\ys) {};
            \node[vs, label={[font=\scriptsize]below:$7$}] (w7) at ({2.5*\xs},-\ys) {};
            \node[vs, label={[font=\scriptsize]above:$2$}] (w2) at ({-2.5*\xs},\ys) {};
            \node[vssel, label={[font=\scriptsize]above:$4$}] (w4) at ({-0.5*\xs},\ys) {};
            \node[vs, label={[font=\scriptsize]above:$6$}] (w6) at ({1.5*\xs},\ys) {};
            \node[vssel, label={[font=\scriptsize]above:$8$}] (w8) at ({3.5*\xs},\ys) {};
            \foreach \i in {1,...,7} {
                    \pgfmathsetmacro{\inext}{int(\i+1)}
                    \draw[eg] (w\i) -- (w\inext);
                }
            \foreach \i in {1,...,6} {
                    \pgfmathsetmacro{\inext}{int(\i+2)}
                    \draw[eg] (w\i) -- (w\inext);
                }
            \node at (0,-1.2) {(b) Zigzag ladder};
        \end{scope}
    \end{tikzpicture}
    \caption{Examples of ECF frustration graphs.
        (a)~Path graph corresponding to the Ising/XY chain, in which vertices $j$ and $k$ are adjacent when $|j-k|=1$.
        (b)~Zigzag ladder corresponding to the Fendley model~\eqref{eq:ffd-rep} for $M=8$, in which vertices $j$ and $k$ are adjacent when $|j-k| \leq 2$.
        The shaded vertices provide examples of independent sets, with $|S|=2$ in (a) and $|S|=3$ in (b).}
    \label{fig:ecf-examples}
\end{figure}

For example, this general construction also covers the Fendley model~\cite{FFD-fendley-2019} with dissipation on a chain of $M$ sites.
Its Hamiltonian is $H=\sum_{j=1}^{M} b_j h_j$, where the $b_j$ are arbitrary coupling constants and
\begin{align}
    \label{eq:ffd-rep}
    h_j &= \sigma_{j-2}^z \sigma_{j-1}^z \sigma_j^x
    \qquad (j=3,\ldots,M)\,,
\end{align}
while $h_1=\sigma_1^x$ and $h_2=\sigma_1^z \sigma_2^x$~\footnote{Fendley's original model is defined on an $(M+2)$ site spin chain with local terms $h_j=\sigma_j^z \sigma_{j+1}^z \sigma_{j+2}^x$ for $j=1,\ldots,M$, so that $H=\sum_{j=1}^{M} b_j h_j$.
The difference amounts only to a trivial degeneracy associated with the first two spins and does not affect the essential structure of the model.}.
Here $\sigma_j^x$ and $\sigma_j^z$ denote the standard Pauli operators acting on site $j$.
As explained below, the corresponding frustration graph has vertex set $V(G)=\{1,\ldots,M\}$ and is the zigzag ladder graph shown in Fig.~\ref{fig:ecf-examples}(b).
One exactly solvable choice of dissipation is the single right-boundary jump operator $\ell = \sqrt{\gamma}\,\sigma_M^z$.
We first review the general FFD framework, within which this model will be identified as a particular case.

\secc{Free fermions in disguise}
We briefly review the FFD framework~\cite{FFD-fendley-2019,FFD-chapman-2021,FFD-chapman-unified-2023}.
We first introduce a Hamiltonian represented by a graph-Clifford algebra defined on a frustration graph $G$:
\begin{equation}
    \label{eq:H-general}
    H_{G} = \sum_{\jbm \in V(G)} b_{\jbm} h_{\jbm},
\end{equation}
where $V(G)$ denotes the vertex set of $G$, $b_{\jbm}$ are arbitrary coupling constants, and $h_{\jbm}$ are generators associated with each vertex $\jbm$ satisfying
\begin{equation}
    \label{eq:graph-clifford}
    (h_{\jbm})^2 = \mathbb{1}, \quad
    h_{\jbm} h_{\ellbm} = (-1)^{A_{\jbm\ellbm}} h_{\ellbm} h_{\jbm},
\end{equation}
where $A$ is the adjacency matrix of $G$ with $A_{\jbm\ellbm} \in \{0,1\}$ for $\jbm \neq \ellbm$ and $A_{\jbm\jbm} = 0$.
Thus, generators associated with adjacent vertices anticommute, while those associated with non-adjacent vertices commute.
Any graph-Clifford algebra~\eqref{eq:graph-clifford} naturally admits a Pauli-string representation on $|V(G)|$ spin-$1/2$ sites; an explicit construction is given in End Matter~\cite{fukai-vona-pozsgay-ffd-eigenstates-paper1-2026}.

\begin{figure}[tb]
    \centering
    \begin{tikzpicture}[
            vs/.style={circle, draw=black, minimum size=2mm, inner sep=0.5mm, line width=0.4mm},
            eg/.style={draw=black, line width=0.5pt}
        ]
        \begin{scope}[shift={(-2.4,0)}]
            \node[vs] (a) at (90:0.7) {};
            \node[vs] (b) at (210:0.7) {};
            \node[vs] (c) at (330:0.7) {};
            \node[vs] (center) at (0,0) {};
            \draw[eg] (center) -- (a);
            \draw[eg] (center) -- (b);
            \draw[eg] (center) -- (c);
            \node at (0,-1.4) {(a) Claw $K_{1,3}$};
        \end{scope}
        \draw[dotted, gray, line width=0.5pt] (-1.1,-1.1) -- (-1.1,1.1);
        \begin{scope}[shift={(0.3,0)}]
            \foreach \i in {1,...,4} {
                    \node[vs] (u\i) at ({45+90*(\i-1)}:0.55) {};
                }
            \foreach \i in {1,...,3} {
                    \pgfmathsetmacro{\inext}{int(\i+1)}
                    \draw[eg] (u\i) -- (u\inext);
                }
            \draw[eg] (u4) -- (u1);
            \node at (0,0) {\small $C_4$};
        \end{scope}
        \begin{scope}[shift={(2.1,0)}]
            \foreach \i in {1,...,6} {
                    \node[vs] (v\i) at ({90+60*(\i-1)}:0.7) {};
                }
            \foreach \i in {1,...,5} {
                    \pgfmathsetmacro{\inext}{int(\i+1)}
                    \draw[eg] (v\i) -- (v\inext);
                }
            \draw[eg] (v6) -- (v1);
            \node at (0,0) {\small $C_6$};
        \end{scope}
        \node at (3.4,0) {$\cdots$};
        \node at (1.5,-1.4) {(b) Even holes};
    \end{tikzpicture}
    \caption{Forbidden induced subgraphs for the ECF condition; a subgraph is \emph{induced} if it contains all edges of the original graph between its vertices.
        (a)~The claw $K_{1,3}$: a center vertex connected to three mutually non-adjacent leaves.
        (b)~Even holes $C_4, C_6, \dots$: induced cycles of even length.}
    \label{fig:claw-evenhole}
\end{figure}
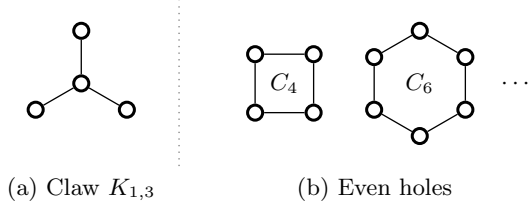

The solvability of $H_G$ is governed by the graph-theoretic properties of $G$.
A subgraph of $G$ is \emph{induced} if, for every pair of its vertices, the two vertices are adjacent in the subgraph exactly when they are adjacent in $G$.
A graph is \emph{claw-free} if it has no claw as an induced subgraph (Fig.~\ref{fig:claw-evenhole}(a)), and \emph{even-hole-free} if it has no even hole as an induced subgraph (Fig.~\ref{fig:claw-evenhole}(b)).
A graph satisfying both conditions is called \emph{even-hole-free, claw-free} (ECF), whose examples are shown in Fig.~\ref{fig:ecf-examples}.
The path graph~(a) corresponds to the Ising/XY chain, solvable by the standard Jordan--Wigner transformation~\cite{jordan1993,ogura-free-fermion-graph-2020,chapman2020characterization}.
The zigzag ladder graph~(b), which corresponds to the Fendley model~\eqref{eq:ffd-rep}, represents a new class of exactly solvable models beyond the Jordan--Wigner paradigm~\cite{FFD-fendley-2019,fendley-pozsgay-2024}.

When $G$ is ECF, the Hamiltonian~\eqref{eq:H-general} is exactly diagonalized~\cite{FFD-fendley-2019,FFD-chapman-2021}:
\begin{equation}
    \label{eq:H-diagonalized}
    H_G = \sum_{k=1}^{\alpha_G} \varepsilon_k \qty[\Psi_k, \Psi_{-k}],
\end{equation}
where $\alpha_G$ is the independence number of $G$ explained below, $\Psi_{\pm k}$ are hidden fermion operators satisfying $\{\Psi_k,\Psi_{k'}\} = \delta_{k+k', 0}$, and $\varepsilon_k$ is the single-particle energy.
The eigenenergies are given by $2\sum_{k} \varepsilon_k s_k + E_0$ with $s_k \in \{0,1\}$, where $E_0 \equiv -\sum_{k=1}^{\alpha_G} \varepsilon_k$ is the ground-state energy.

The single-particle energies $\varepsilon_k$ are determined by the \emph{independence polynomial} of $G$.
An \emph{independent set} $S \subseteq V(G)$ is a subset of mutually non-adjacent vertices, so that the corresponding generators all commute.
For example, the shaded vertices in Fig.~\ref{fig:ecf-examples} form independent sets: $S=\{2,4\}$ for the path graph in (a) and $S=\{1,4,8\}$ for the zigzag ladder in (b), the frustration graph of the Fendley model~\eqref{eq:ffd-rep}.
The \emph{independence number} $\alpha_G$ is the maximum size of an independent set.
The independence polynomial is defined as
\begin{equation}
    \label{eq:independence-polynomial}
    P_G(x) \equiv \sum_{S \in \indepset{G}} (-x)^{|S|} \prod_{\jbm \in S} b_{\jbm}^2,
\end{equation}
where $\indepset{G}$ denotes the collection of all independent sets.
The single-particle energies are given by $\varepsilon_k = 1 / u_k$, where $u_1, \dots, u_{\alpha_G}$ are the roots of $P_G(u^2)=0$.
For claw-free graphs, if all $b_{\jbm}$ are real, so that the vertex weights $b_{\jbm}^2$ are non-negative, then $P_G(x)$ has only positive real roots~\footnote{The corresponding real-rootedness result was first proved for the unweighted independence polynomial~\cite{chudnovsky-seymour}; see also Ref.~\cite{bencs-independence-polynomial}.
References~\cite{engstrom-weighted,leake-ryder-independence} extend this real-rootedness result to weighted independence polynomials and define the independence polynomial as $\sum_{S} x^{|S|} \prod b_{\jbm}^2$ without the $(-1)^{|S|}$ sign, so that the roots are real \emph{negative}.
Our convention with $(-x)^{|S|}$ absorbs this sign, making the roots positive.}.
Therefore the single-particle energies are positive.

The free-fermion operators in Eq.~\eqref{eq:H-diagonalized} are built from an \emph{edge operator} $\chi$ and a transfer matrix $T_G(u)$~\cite{FFD-fendley-2019,FFD-chapman-2021,FFD-chapman-unified-2023}:
\begin{align}
    \Psi_{\pm k}
    &=
    \frac{1}{\mathcal{N}_k} T_G(\mp u_k) \chi T_G(\pm u_k),
    \label{eq:free-fermion-mode}
\end{align}
where $u_k$ is again a root of $P_G(u^2)=0$.
The normalization constant $\mathcal{N}_k$ is given in End Matter [see Eq.~\eqref{eq:mode-normalization}].
The edge operator $\chi$ and the transfer matrix $T_G(u)$ are defined below.

\begin{figure}[tb]
    \centering
    \begin{tikzpicture}[
            x=0.85cm, y=0.85cm,
            v/.style={circle, draw=black, fill=white, minimum size=2mm, inner sep=0.5mm, line width=0.4mm},
            ks/.style={v, fill=orange!35},
            xv/.style={v, fill=ForestGreen!20},
            eg/.style={draw=black, line width=0.5pt},
            good/.style={draw=ForestGreen!70!black, line width=0.95pt},
            bad/.style={draw=BrickRed, line width=0.95pt}
        ]
        \begin{scope}[shift={(-2.3,0)}]
            \node[ks, label=left:$A$] (u1) at (0,0.55) {};
            \node[ks, label=left:$B$] (v1) at (0,-0.55) {};
            \node[xv, label=left:$\chi$] (chi1) at (-1.15,0) {};
            \node[v, label=above:$C$] (x1) at (1.25,0.98) {};
            \node[v, label=below:$D$] (y1) at (1.55,-0.28) {};

            \draw[eg] (u1)--(v1);
            \draw[eg] (chi1)--(u1);
            \draw[eg] (chi1)--(v1);
            \draw[eg] (u1)--(x1);
            \draw[eg] (u1)--(y1);
            \draw[eg] (v1)--(y1);
            \draw[good] (x1)--(y1);

            \node[font=\scriptsize] at (0,-1.7) {(a) Simplicial};
        \end{scope}

        \begin{scope}[shift={(2.3,0)}]
            \node[ks, label=left:$A$] (u2) at (0,0.55) {};
            \node[ks, label=left:$B$] (v2) at (0,-0.55) {};
            \node[xv, label=left:$\chi$] (chi2) at (-1.15,0) {};
            \node[v, label=above:$C$] (x2) at (1.25,0.98) {};
            \node[v, label=below:$D$] (y2) at (1.55,-0.28) {};

            \draw[eg] (u2)--(v2);
            \draw[eg] (chi2)--(v2);
            \draw[eg] (v2)--(y2);
            \draw[bad] (chi2)--(u2);
            \draw[bad] (u2)--(x2);
            \draw[bad] (u2)--(y2);

            \node[font=\scriptsize] at (0,-1.7) {(b) Not simplicial};
        \end{scope}
    \end{tikzpicture}
    \caption{
        Simplicial vs.\ non-simplicial cliques.
        The claw-free graph $G$ consists of four vertices $A,B,C,D$, and $G_\chi$ is obtained by adding the vertex $\chi$ (green) connected to the clique $K_s = \{A,B\}$ (orange).
        (a) $G_\chi$ is also claw-free, so $K_s$ is simplicial.
        (b) Without the edge $C\md D$, $\Gamma[A]\setminus K_s = \{C,D\}$ is not a clique, so $K_s$ is not simplicial; indeed, adding $\chi$ creates a claw centered at $A$ whose leaves are drawn in red.
    }
    \label{fig:simplicial-clique}
\end{figure}
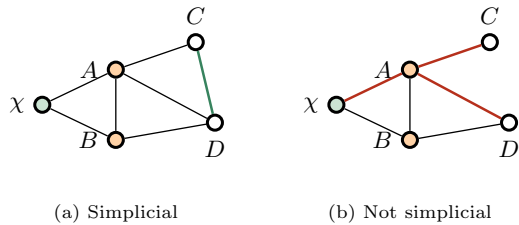

The edge operator $\chi$ is associated with a clique $K_s \subseteq V(G)$, i.e., a subset of mutually adjacent vertices.
It satisfies $\chi^2 = 1$ and
\begin{equation}
    \label{eq:edge-operator}
    \{h_{\jbm}, \chi\} = 0 \;\text{ for } \jbm \in K_s, \quad [h_{\jbm}, \chi] = 0 \;\text{ for } \jbm \notin K_s.
\end{equation}
In the frustration graph, this corresponds to adding a new vertex $\chi$ adjacent to every vertex of $K_s$ and to no others, giving an \emph{extended graph} $G_\chi$.
The clique $K_s$ is called \emph{simplicial} if $G_\chi$ is also claw-free~\cite{evenhole-simplicial,FFD-chapman-unified-2023}.
Equivalently, denoting the closed neighborhood of $\jbm$ by $\Gamma[\jbm] \equiv\{\jbm\} \cup \qty{\ellbm \in V(G) \mid A_{\jbm\ellbm}=1}$, $K_s$ is simplicial if and only if
$
    \Gamma[\jbm] \setminus K_s \quad \text{is a clique for all } \jbm \in K_s.
$
Figure~\ref{fig:simplicial-clique} shows (a) a simplicial clique and (b) a non-simplicial clique.
The edge operator $\chi$ can be represented as a Pauli string~\cite{fukai-vona-pozsgay-ffd-eigenstates-paper1-2026}; see Eq.~\eqref{eq:defining-representation-edge} in End Matter.

In fact, the even-hole-free condition can be relaxed: any claw-free frustration graph with a simplicial clique yields hidden free-fermion solvability~\cite{FFD-chapman-unified-2023}, and the ECF is a special case~\cite{FFD-fendley-2019,FFD-chapman-2021}.

The transfer matrix $T_G(u)$ is defined as a generating function of the conserved charges $Q_{G}^{(n)}$ associated with independent sets of size $n$:
\begin{equation}
    Q_{G}^{(n)} \equiv \sum_{\substack{S \in \indepset{G}\\ |S|=n}} \prod_{\jbm \in S} b_{\jbm} h_{\jbm},
    \quad
    T_G(u) \equiv \sum_{n=0}^{\alpha_G} (-u)^n Q_{G}^{(n)},
\end{equation}
with $Q_{G}^{(0)} \equiv \mathbb{1}$.
It satisfies $T_G(u)T_G(-u)=P_G(u^2)$.
For claw-free graphs, these charges commute, so $[T_G(u),T_G(v)] = 0$~\cite{FFD-fendley-2019,FFD-chapman-2021,FFD-chapman-unified-2023}.

In this work, we extend the FFD framework to dissipative systems.
As we show below, $\chi$ also plays a central role in the dissipative setting.

\secc{FFD solvable Liouvillian}
The graph-theoretic solvability criterion extends to open quantum systems.
If the Hamiltonian and jump operators are designed so that the frustration graph $\liouvillian{G}$ of the non-Hermitian Hamiltonian $\Hi$ in Eq.~\eqref{eq:liouvillian-doubled} is claw-free and has a simplicial clique, $\Hi$ is solvable by hidden free fermions~\cite{FFD-chapman-unified-2023}.

For an ECF graph $G$, choose an edge operator $\chi$ associated with a simplicial clique $K_s \subseteq V(G)$, which can be chosen Hermitian; see Eq.~\eqref{eq:defining-representation-edge} in End Matter.
Take a single jump operator
\begin{align}
    \ell = \sqrt{\gamma}\,\chi.
\end{align}
The resulting $\liouvillian{G}$ satisfies the solvability criterion above.
The corresponding non-Hermitian Hamiltonian on the doubled space in Eq.~\eqref{eq:liouvillian-doubled} becomes
\begin{align}
    \label{eq:ffd-liouvillian-general}
    \Hi_{\liouvillian{G}}
    =
    \sum_{\jbm \in V(G)} b_{\jbm}\qty(h_{\jbm}^{(1)} - h_{\jbm}^{(2)})
    + \imi\gamma d - \imi\gamma,
\end{align}
where we defined $d \equiv \chi \otimes \chi$.
The operators $\{h_{\jbm}^{(1)}\}$ and $\{h_{\jbm}^{(2)}\}$ form two independent copies of the original graph-Clifford algebra, and $d$ satisfies $d^2 = \mathbb{1}$ and
\begin{align}
    \{d, h_{\jbm}^{(p)}\} = 0 \ (\jbm \in K_s)\,,\quad
    [d, h_{\jbm}^{(p)}]   = 0 \ (\jbm \notin K_s)\,,
\end{align}
where $p=1,2$ labels the two copies.
The Liouvillian frustration graph $\liouvillian{G}$ consists of two disjoint copies $G^{(1)}$ and $G^{(2)}$ of $G$, together with one extra vertex $d$ connected to all vertices in $K_s^{(1)} \cup K_s^{(2)}$.
If $G$ is ECF and $K_s$ is simplicial, then $\liouvillian{G}$ is again ECF: even holes cannot span both copies, and claws involving $d$ are excluded by the simpliciality of $K_s$.
For $p=1,2$, define
\begin{align}
    \liouvillian{K}_s^{(p)} \equiv \{d\} \cup K_s^{(p)}.
\end{align}
These are simplicial cliques of $\liouvillian{G}$.
We denote by $\liouvillian{\chi}$ the edge operator of $\liouvillian{G}$ associated with $\liouvillian{K}_s^{(1)}$.
Therefore the FFD construction of Refs.~\cite{FFD-fendley-2019,FFD-chapman-2021,FFD-chapman-unified-2023} applies also to the non-Hermitian Hamiltonian $\Hi_{\liouvillian{G}}$.

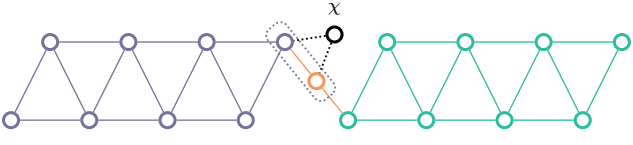
\begin{figure}[tb]
    \centering
    \begin{tikzpicture}[
            vsleft/.style={circle, draw=CadetBlue, minimum size=2mm, inner sep=0.5mm, line width=0.4mm},
            vsright/.style={circle, draw=SeaGreen, minimum size=2mm, inner sep=0.5mm, line width=0.4mm},
            vcenter/.style={circle, draw=Peach, minimum size=2mm, inner sep=0.5mm, line width=0.4mm},
            vschi/.style={circle, draw=black, minimum size=2mm, inner sep=0.5mm, line width=0.4mm},
            edgeleft/.style={draw=CadetBlue, line width=0.5pt},
            edgeright/.style={draw=SeaGreen, line width=0.5pt},
            edgecenter/.style={draw=Peach, line width=0.5pt},
            cliquemark/.style={draw=CadetBlue, densely dotted, line width=0.7pt, rounded corners=3pt},
            chiedge/.style={draw=black, densely dotted, line width=0.7pt}
        ]
        \pgfmathsetmacro{\Mval}{8}  
        \pgfmathsetmacro{\xscale}{0.52}
        \pgfmathsetmacro{\yscale}{0.6}
        \pgfmathsetmacro{\gap}{-0.1}  

        \pgfmathsetmacro{\nodd}{int((\Mval+1)/2)}
        \pgfmathsetmacro{\neven}{int(\Mval/2)}
        \pgfmathsetmacro{\yrowone}{1.732*\yscale/2}
        \pgfmathsetmacro{\yrowtwo}{-1.732*\yscale/2}
        \pgfmathsetmacro{\xoffset}{\gap + \Mval*\xscale}  

        \node[vcenter] (center) at (0, 0) {};

        \foreach \i in {1,...,\nodd} {
                \pgfmathsetmacro{\v}{int(2*\i-1)}
                \pgfmathsetmacro{\xpos}{-\gap - (\Mval - \v + 1)*\xscale}
                \node[vsleft] (L\v) at (\xpos, -\yrowone) {};
            }
        \foreach \i in {1,...,\neven} {
                \pgfmathsetmacro{\v}{int(2*\i)}
                \pgfmathsetmacro{\xpos}{-\gap - (\Mval - \v + 1)*\xscale}
                \node[vsleft] (L\v) at (\xpos, -\yrowtwo) {};
            }
        \pgfmathsetmacro{\edgeTwo}{int(\Mval-2)}
        \foreach \i in {1,...,\edgeTwo} {
                \pgfmathsetmacro{\inext}{int(\i+2)}
                \draw[edgeleft] (L\i) -- (L\inext);
            }
        \pgfmathsetmacro{\edgeOne}{int(\Mval-1)}
        \foreach \i in {1,...,\edgeOne} {
                \pgfmathsetmacro{\inext}{int(\i+1)}
                \draw[edgeleft] (L\i) -- (L\inext);
            }

        \foreach \i in {1,...,\nodd} {
                \pgfmathsetmacro{\v}{int(2*\i-1)}
                \pgfmathsetmacro{\xpos}{\gap + (\Mval - \v + 1)*\xscale}
                \node[vsright] (R\v) at (\xpos, \yrowone) {};
            }
        \foreach \i in {1,...,\neven} {
                \pgfmathsetmacro{\v}{int(2*\i)}
                \pgfmathsetmacro{\xpos}{\gap + (\Mval - \v + 1)*\xscale}
                \node[vsright] (R\v) at (\xpos, \yrowtwo) {};
            }
        \foreach \i in {1,...,\edgeTwo} {
                \pgfmathsetmacro{\inext}{int(\i+2)}
                \draw[edgeright] (R\i) -- (R\inext);
            }
        \foreach \i in {1,...,\edgeOne} {
                \pgfmathsetmacro{\inext}{int(\i+1)}
                \draw[edgeright] (R\i) -- (R\inext);
            }

        \pgfmathsetmacro{\connA}{int(\Mval)}
        \pgfmathsetmacro{\connB}{int(\Mval-1)}
        \pgfmathsetmacro{\connC}{int(\Mval-2)}
        \draw[edgecenter] (center) -- (L\connA);
        \draw[edgecenter] (center) -- (R\connA);

        \coordinate (cliqueMid) at ($(center)!0.5!(L\connA)$);
        \node[vschi] (chiL) at ($(center)+(0.24,0.62)$) {};
        \node[text=black, font=\scriptsize, above=1pt of chiL] {$\liouvillian{\chi}$};
        \draw[chiedge] (chiL) -- (center);
        \draw[chiedge] (chiL) -- (L\connA);
        \begin{scope}[rotate around={129:(cliqueMid)}]
            \draw[cliquemark] ($(cliqueMid)+(-0.6,-0.16)$) rectangle ($(cliqueMid)+(0.6,0.16)$);
        \end{scope}

    \end{tikzpicture}
    \caption{Liouvillian frustration graph $\liouvillian{G}_M$ for the boundary-driven Fendley model.
        The left (blue) and right (green) subgraphs are the two copies $G^{(1)}$ and $G^{(2)}$ of the zigzag ladder in Fig.~\ref{fig:ecf-examples}(b).
        The central orange vertex is $d=\chi\otimes\chi=\sigma_M^{z}\otimes \sigma_M^{z}$, shown here for the single-vertex boundary clique $K_s=\{M\}$.
        The extra blue vertex represents the edge operator $\liouvillian{\chi}$, and the dotted contour marks the simplicial clique $\liouvillian{K}_s^{(1)}=\{d\}\cup K_s^{(1)}$.}
    \label{fig:Li-FFD-single-g}
\end{figure}

As a concrete example, for the Fendley model with dissipation introduced above, taking $K_s=\{M\}$ and $\chi=\sigma_M^z$ yields the Liouvillian frustration graph $\liouvillian{G}_M$ shown in Fig.~\ref{fig:Li-FFD-single-g} (ignore the dotted lines for the moment).

The independence polynomial of $\liouvillian{G}$ splits according to whether an independent set contains $d$:
\begin{align}
    \poly{\liouvillian{G}}{u^2}
     & =
    P_G(u^2)^2 + \gamma^2 u^2 P_{G \setminus K_s}(u^2)^2
    \nonumber \\
     & =
    \polytilpm{G}{+}{u} \polytilpm{G}{-}{u},
    \label{eq:Li-general-factorized}
\end{align}
where the first term counts independent sets excluding $d$, and the second those including $d$, and we define
\begin{align}
    \polytilpm{G}{\pm}{u}
    \equiv
    P_G(u^2) \pm \imi\gamma u P_{G \setminus K_s}(u^2)\,.
    \label{eq:Li-general-Ppm}
\end{align}
Because the added vertex $d$ carries the imaginary weight $\imi\gamma$, the usual real-rootedness result for weighted independence polynomials of claw-free graphs does not apply here~\cite{engstrom-weighted,leake-ryder-independence}, so the corresponding single-particle energies can be complex.

We next define the free-fermion modes of $\Hi_{\liouvillian{G}}$ using the FFD construction:
\begin{align}
    \Psitil_{\pm k}
    =
    \frac{1}{\Ntil_k} T_{\liouvillian{G}}(\mp \liouvillian{u}_{k})\liouvillian{\chi}T_{\liouvillian{G}}(\pm \liouvillian{u}_{k}),
\end{align}
where $\polytilpm{G}{+}{\liouvillian{u}_k}=0$ for $k=1,\dots,\alpha_{\liouvillian{G}}$, $\Ntil_k$ is a normalization factor whose explicit form is given in End Matter, Eq.~\eqref{eq:liouvillian-mode-normalization}, and $\alpha_{\liouvillian{G}}=\max\qty(2\alpha_G,\;1+2\alpha_{G \setminus K_s})$.
Both $\polytilpm{G}{\pm}{u}$ are degree-$\alpha_{\liouvillian{G}}$ polynomials in $u$, related by $\polytilpm{G}{-}{u}=\polytilpm{G}{+}{-u}$ and $\polytilpm{G}{\pm}{u}=\overline{\polytilpm{G}{\pm}{-\overline{u}}}$.
We expect $\Im \liouvillian{u}_k>0$ to hold for general ECF graphs, as observed numerically in the boundary-driven Fendley model, so that $\liouvillian{\varepsilon}_k \equiv 1/\liouvillian{u}_k$ satisfies $\Im \liouvillian{\varepsilon}_k<0$.
We take $\Psitil_{-k}$ as the annihilation operator and $\Psitil_{k}$ as the creation operator for the mode with single-particle energy $\liouvillian{\varepsilon}_k$.
Because $\Hi_{\liouvillian{G}}$ is non-Hermitian, $\Psitil_{k}$ is not in general the Hermitian adjoint of $\Psitil_{-k}$, namely $\Psitil_k^\dagger \neq \Psitil_{-k}$.
However, the fermionic anticommutation relations still hold: $\{\Psitil_{k}, \Psitil_{k'}\} = \delta_{k+k', 0}$~\cite{Ulrich-Bilstein-1999, katsura-lindblad-1}.

Then, the Liouvillian~\eqref{eq:ffd-liouvillian-general} is diagonalized as
\begin{align}
    \Hi_{\liouvillian{G}}
     =
     \sum_{k = 1}^{\alpha_{\liouvillian{G}}} \liouvillian{\varepsilon}_k \qty[\Psitil_{k}, \Psitil_{-k}] - \imi\gamma
     =
    2 \sum_{k=1}^{\alpha_{\liouvillian{G}}} \liouvillian{\varepsilon}_k \liouvillian{n}_k,
    \label{eq:Li-general-diag}
\end{align}
where $\liouvillian{n}_k \equiv \Psitil_{k} \Psitil_{-k}$ is the occupation number of the mode $k$.
Comparing the coefficient of $u$ in Eq.~\eqref{eq:Li-general-Ppm} with that in $\polytilpm{G}{+}{u}=\prod_k(1-u/\liouvillian{u}_k)$ gives $\sum_k \liouvillian{\varepsilon}_k=-\imi\gamma$.
Using $[\Psitil_{k}, \Psitil_{-k}]=2\liouvillian{n}_k-1$, the commutator term contributes the constant $-\sum_k \liouvillian{\varepsilon}_k=\imi\gamma$, which cancels $-\imi\gamma$ in the middle expression of Eq.~\eqref{eq:Li-general-diag} and yields the right-hand side.
The configuration with $\liouvillian{n}_k=0$ for all $k$ therefore gives the zero Liouvillian eigenvalue.
Each occupied mode then contributes $-2\imi\liouvillian{\varepsilon}_k$ to the Liouvillian eigenvalue, so $\lambda=-2\imi \sum_k \liouvillian{\varepsilon}_k s_k$ for $s_k\in\{0,1\}$, satisfying $\Re \lambda \le 0$.
For generic couplings~\footnote{At fine-tuned values of $b_j$, additional accidental degeneracies may occur.
Throughout this work, ``generic couplings'' means that such nongeneric degeneracies are excluded.}, these $2^{\alpha_{\liouvillian{G}}}$ Liouvillian eigenenergies are distinct.

In addition to the $\alpha_{\liouvillian{G}}$ hidden fermion modes $\Psitil_{\pm k}$, there exist zero modes $\Xi_j$ satisfying $\Xi_j^2=1$, $[\Xi_j, \Hi_{\liouvillian{G}}]=0$, $\Xi_l \Xi_k = \pm \Xi_k \Xi_l$, and $\Psitil_k \Xi_l = \pm \Xi_l \Psitil_k$~\cite{Istvan-ffd-prb-2025,Eric-FFD-hilbert-space-2025}.
These generate a $2^{2\abs{V(G)}-\alpha_{\liouvillian{G}}}$-fold degeneracy at each eigenenergy.
Accordingly, the zero-eigenvalue eigenspace $\Ker \Li$ is $2^{2\abs{V(G)}-\alpha_{\liouvillian{G}}}$-dimensional.
Physical steady states are the positive, trace-one operators in this eigenspace.

\secc{Dispersion relation}
Since $\polytilpm{G}{+}{u}$ can be obtained explicitly for any ECF graph, so can the dispersion relation.
Using $\lambda=-2\imi \sum_k \liouvillian{\varepsilon}_k s_k$ and $\Im \liouvillian{\varepsilon}_k \le 0$, the Liouvillian gap is given by
\begin{align}
    g
    \equiv
    -\max_{\lambda\neq 0}\Re \lambda
    =
    -2 \max_{1 \le k \le \alpha_{\liouvillian{G}}} \Im \liouvillian{\varepsilon}_k\,.
\end{align}

For the boundary-driven Fendley model [Eq.~\eqref{eq:ffd-rep}, Fig.~\ref{fig:Li-FFD-single-g}] with uniform couplings $b_j=1$ and $M=3(L-1)$, the Liouvillian gap is given by
\begin{align}
    g
    =
    \frac{\gamma\pi^2}{3+\gamma^2}\frac{1}{L^3}\,.
    \label{eq:uniform-L3}
\end{align}
The standing-wave analysis leading to Eq.~\eqref{eq:uniform-L3} is omitted here.
This $L^{-3}$ scaling is consistent with that observed in other integrable chains with boundary dissipation~\cite{third-quantization,prosen-pizorn-2008,medvedyeva-kehrein-2014,znidaric-pre-2015,shibata-boundary-dephasing-2020}.
Note that the isolated Fendley model has an energy gap scaling as $L^{-3/2}$~\cite{FFD-fendley-2019}.
For inhomogeneous couplings, a phase with an exponentially small Liouvillian gap also appears~\cite{znidaric-pre-2015}.

\secc{Infinite-temperature autocorrelation function}
As a representative case, we consider the infinite-temperature autocorrelation function of the edge operator $\chi$,
\begin{align}
    B(t) \equiv \ev{\chi(t)\chi(0)}_{\infty},
\end{align}
with $\chi(t)=e^{t\Li^\dagger}[\chi]$ and $\ev{\cdot}_{\infty} \equiv \tr(\cdot)/\tr(\mathbb{1})$, where $\Li^\dagger$ is the adjoint of the Liouvillian given in Eq.~\eqref{eq:Li-adjoint}.
Extending the Krylov-space method of Refs.~\cite{Istvan-ffd-prb-2025,fukai-vona-pozsgay-ffd-eigenstates-paper2-2026} to the boundary-driven Liouvillian (see End Matter for the derivation), we obtain
\begin{align}
    B(t)
    =
    -\sum_{k=1}^{\alpha_{\liouvillian{G}}}
    \frac{P_{G \setminus K_s}(\liouvillian{u}_k^2)}
    {\liouvillian{u}_k\,\polytilpm{G}{+\prime}{\liouvillian{u}_k}}
    \exp\!\qty[-2 \qty(\gamma - \imi \liouvillian{\varepsilon}_k) t],
    \label{eq:Bt-residue}
\end{align}
where $\polytilpm{G}{+\prime}{u}\equiv\partial_u\polytilpm{G}{+}{u}$.

For the uniform boundary-driven Fendley model, the long-time asymptotics in the thermodynamic limit is
\begin{align}
    B(t)
    \sim
    \frac{\gamma^2(\gamma^2+3)}{(1+\gamma^2)^2}\exp\qty[-\frac{2\gamma}{1+\gamma^2}t],
    \quad
    t\gg 1.
    \label{eq:Bt-uniform-asymptotic}
\end{align}
The relaxation rate $\frac{2\gamma}{1+\gamma^2}$ increases monotonically for $\gamma<1$.
However, for $\gamma>1$, it decreases monotonically with $\gamma$, which is a typical signature of the continuous Quantum Zeno effect~\cite{syassen-StrongDissipationInhibits-2008,yan-ObservationDipolarSpinexchange-2013}.
We also note that, in the closed-system limit $\gamma=0$, this exponential decay is replaced by the algebraic tail $B(t)\sim t^{-2/3}$~\cite{Istvan-ffd-prb-2025}.

\secc{Conclusion}
By extending the FFD solvability to open quantum systems, we established a new class of exactly solvable Lindbladians that are not reducible to quadratic forms in Jordan--Wigner fermions, distinct from the third-quantization approach to quadratic open systems~\cite{third-quantization,third-quantization-2}.
For generic couplings, any even-hole-free, claw-free frustration graph yields an exactly solvable Lindbladian once an edge operator associated with a simplicial clique is taken as the jump operator.
The exact solution~\eqref{eq:Li-general-diag} enables the calculation of the Liouvillian gap and yields a closed-form expression for the infinite-temperature autocorrelation function~\eqref{eq:Bt-residue}.

More generally, the Liouvillian remains free-fermion solvable whenever its frustration graph is claw-free and has a simplicial clique~\cite{FFD-chapman-unified-2023}.
For example, adding a second jump operator at the opposite edge of the boundary-driven Fendley model creates even holes in the Liouvillian frustration graph, but the resulting graph remains claw-free and has a simplicial clique, so the Liouvillian is still free-fermion solvable~\footnote{ECF is a special case because every even-hole-free claw-free graph has a simplicial clique~\cite{evenhole-simplicial}.}.
The same criterion allows exactly solvable dissipative extensions of other FFD models whose frustration graphs are claw-free and have a simplicial clique, including the Kitaev honeycomb model~\cite{kitaev2006anyons}.

\begin{acknowledgments}
    K.F. was supported by JSPS KAKENHI Grant No.~JP25K23354.
    H.K. was supported by JSPS KAKENHI Grants No.~JP23K25783 and No.~JP23K25790.
    K.F. and H.K. were supported by MEXT KAKENHI Grant-in-Aid for Transformative Research Areas A ``Extreme Universe" (KAKENHI Grant No.~JP21H05191).
\end{acknowledgments}


\bibliography{ref}

\begin{thebibliography}{80}%
\makeatletter
\providecommand \@ifxundefined [1]{%
 \@ifx{#1\undefined}
}%
\providecommand \@ifnum [1]{%
 \ifnum #1\expandafter \@firstoftwo
 \else \expandafter \@secondoftwo
 \fi
}%
\providecommand \@ifx [1]{%
 \ifx #1\expandafter \@firstoftwo
 \else \expandafter \@secondoftwo
 \fi
}%
\providecommand \natexlab [1]{#1}%
\providecommand \enquote  [1]{``#1''}%
\providecommand \bibnamefont  [1]{#1}%
\providecommand \bibfnamefont [1]{#1}%
\providecommand \citenamefont [1]{#1}%
\providecommand \href@noop [0]{\@secondoftwo}%
\providecommand \href [0]{\begingroup \@sanitize@url \@href}%
\providecommand \@href[1]{\@@startlink{#1}\@@href}%
\providecommand \@@href[1]{\endgroup#1\@@endlink}%
\providecommand \@sanitize@url [0]{\catcode `\\12\catcode `\$12\catcode
  `\&12\catcode `\#12\catcode `\^12\catcode `\_12\catcode `\%12\relax}%
\providecommand \@@startlink[1]{}%
\providecommand \@@endlink[0]{}%
\providecommand \url  [0]{\begingroup\@sanitize@url \@url }%
\providecommand \@url [1]{\endgroup\@href {#1}{\urlprefix }}%
\providecommand \urlprefix  [0]{URL }%
\providecommand \Eprint [0]{\href }%
\providecommand \doibase [0]{https://doi.org/}%
\providecommand \selectlanguage [0]{\@gobble}%
\providecommand \bibinfo  [0]{\@secondoftwo}%
\providecommand \bibfield  [0]{\@secondoftwo}%
\providecommand \translation [1]{[#1]}%
\providecommand \BibitemOpen [0]{}%
\providecommand \bibitemStop [0]{}%
\providecommand \bibitemNoStop [0]{.\EOS\space}%
\providecommand \EOS [0]{\spacefactor3000\relax}%
\providecommand \BibitemShut  [1]{\csname bibitem#1\endcsname}%
\let\auto@bib@innerbib\@empty
\bibitem [{\citenamefont {Rigol}\ \emph {et~al.}(2007)\citenamefont {Rigol},
  \citenamefont {Dunjko}, \citenamefont {Yurovsky},\ and\ \citenamefont
  {Olshanii}}]{rigol-gge}%
  \BibitemOpen
  \bibfield  {author} {\bibinfo {author} {\bibfnamefont {M.}~\bibnamefont
  {Rigol}}, \bibinfo {author} {\bibfnamefont {V.}~\bibnamefont {Dunjko}},
  \bibinfo {author} {\bibfnamefont {V.}~\bibnamefont {Yurovsky}},\ and\
  \bibinfo {author} {\bibfnamefont {M.}~\bibnamefont {Olshanii}},\ }\bibfield
  {title} {\bibinfo {title} {{Relaxation in a completely integrable many-body
  quantum system: An ab initio study of the dynamics of the highly excited
  states of 1D lattice hard-core bosons}},\ }\href
  {https://doi.org/10.1103/PhysRevLett.98.050405} {\bibfield  {journal}
  {\bibinfo  {journal} {Phys. Rev. Lett.}\ }\textbf {\bibinfo {volume} {98}},\
  \bibinfo {pages} {050405} (\bibinfo {year} {2007})}\BibitemShut {NoStop}%
\bibitem [{\citenamefont {Ilievski}\ \emph {et~al.}(2015)\citenamefont
  {Ilievski}, \citenamefont {De~Nardis}, \citenamefont {Wouters}, \citenamefont
  {Caux}, \citenamefont {Essler},\ and\ \citenamefont {Prosen}}]{JS-CGGE}%
  \BibitemOpen
  \bibfield  {author} {\bibinfo {author} {\bibfnamefont {E.}~\bibnamefont
  {Ilievski}}, \bibinfo {author} {\bibfnamefont {J.}~\bibnamefont {De~Nardis}},
  \bibinfo {author} {\bibfnamefont {B.}~\bibnamefont {Wouters}}, \bibinfo
  {author} {\bibfnamefont {J.-S.}\ \bibnamefont {Caux}}, \bibinfo {author}
  {\bibfnamefont {F.~H.~L.}\ \bibnamefont {Essler}},\ and\ \bibinfo {author}
  {\bibfnamefont {T.}~\bibnamefont {Prosen}},\ }\bibfield  {title} {\bibinfo
  {title} {{Complete Generalized Gibbs Ensembles in an Interacting Theory}},\
  }\href {https://doi.org/10.1103/PhysRevLett.115.157201} {\bibfield  {journal}
  {\bibinfo  {journal} {Phys. Rev. Lett.}\ }\textbf {\bibinfo {volume} {115}},\
  \bibinfo {pages} {157201} (\bibinfo {year} {2015})}\BibitemShut {NoStop}%
\bibitem [{\citenamefont {Castro-Alvaredo}\ \emph {et~al.}(2016)\citenamefont
  {Castro-Alvaredo}, \citenamefont {Doyon},\ and\ \citenamefont
  {Yoshimura}}]{doyon-ghd}%
  \BibitemOpen
  \bibfield  {author} {\bibinfo {author} {\bibfnamefont {O.~A.}\ \bibnamefont
  {Castro-Alvaredo}}, \bibinfo {author} {\bibfnamefont {B.}~\bibnamefont
  {Doyon}},\ and\ \bibinfo {author} {\bibfnamefont {T.}~\bibnamefont
  {Yoshimura}},\ }\bibfield  {title} {\bibinfo {title} {{Emergent Hydrodynamics
  in Integrable Quantum Systems Out of Equilibrium}},\ }\href
  {https://doi.org/10.1103/PhysRevX.6.041065} {\bibfield  {journal} {\bibinfo
  {journal} {Phys. Rev. X}\ }\textbf {\bibinfo {volume} {6}},\ \bibinfo {pages}
  {041065} (\bibinfo {year} {2016})}\BibitemShut {NoStop}%
\bibitem [{\citenamefont {Bertini}\ \emph {et~al.}(2016)\citenamefont
  {Bertini}, \citenamefont {Collura}, \citenamefont {De~Nardis},\ and\
  \citenamefont {Fagotti}}]{jacopo-ghd}%
  \BibitemOpen
  \bibfield  {author} {\bibinfo {author} {\bibfnamefont {B.}~\bibnamefont
  {Bertini}}, \bibinfo {author} {\bibfnamefont {M.}~\bibnamefont {Collura}},
  \bibinfo {author} {\bibfnamefont {J.}~\bibnamefont {De~Nardis}},\ and\
  \bibinfo {author} {\bibfnamefont {M.}~\bibnamefont {Fagotti}},\ }\bibfield
  {title} {\bibinfo {title} {{Transport in Out-of-Equilibrium XXZ Chains: Exact
  Profiles of Charges and Currents}},\ }\href
  {https://doi.org/10.1103/PhysRevLett.117.207201} {\bibfield  {journal}
  {\bibinfo  {journal} {Phys. Rev. Lett.}\ }\textbf {\bibinfo {volume} {117}},\
  \bibinfo {pages} {207201} (\bibinfo {year} {2016})}\BibitemShut {NoStop}%
\bibitem [{\citenamefont {Fendley}(2019)}]{FFD-fendley-2019}%
  \BibitemOpen
  \bibfield  {author} {\bibinfo {author} {\bibfnamefont {P.}~\bibnamefont
  {Fendley}},\ }\bibfield  {title} {\bibinfo {title} {{Free fermions in
  disguise}},\ }\href {https://doi.org/10.1088/1751-8121/ab305d} {\bibfield
  {journal} {\bibinfo  {journal} {J. Phys. A}\ }\textbf {\bibinfo {volume}
  {52}},\ \bibinfo {pages} {335002} (\bibinfo {year} {2019})}\BibitemShut
  {NoStop}%
\bibitem [{\citenamefont {Alcaraz}\ and\ \citenamefont
  {Pimenta}(2020{\natexlab{a}})}]{alcaraz-pimenta-multispin-2020}%
  \BibitemOpen
  \bibfield  {author} {\bibinfo {author} {\bibfnamefont {F.~C.}\ \bibnamefont
  {Alcaraz}}\ and\ \bibinfo {author} {\bibfnamefont {R.~A.}\ \bibnamefont
  {Pimenta}},\ }\bibfield  {title} {\bibinfo {title} {{Free fermionic and
  parafermionic quantum spin chains with multispin interactions}},\ }\href
  {https://doi.org/10.1103/PhysRevB.102.121101} {\bibfield  {journal} {\bibinfo
   {journal} {Phys. Rev. B}\ }\textbf {\bibinfo {volume} {102}},\ \bibinfo
  {pages} {121101} (\bibinfo {year} {2020}{\natexlab{a}})}\BibitemShut
  {NoStop}%
\bibitem [{\citenamefont {Alcaraz}\ and\ \citenamefont
  {Pimenta}(2020{\natexlab{b}})}]{alcaraz-pimenta-spectrum-2020}%
  \BibitemOpen
  \bibfield  {author} {\bibinfo {author} {\bibfnamefont {F.~C.}\ \bibnamefont
  {Alcaraz}}\ and\ \bibinfo {author} {\bibfnamefont {R.~A.}\ \bibnamefont
  {Pimenta}},\ }\bibfield  {title} {\bibinfo {title} {{Integrable quantum spin
  chains with free fermionic and parafermionic spectrum}},\ }\href
  {https://doi.org/10.1103/PhysRevB.102.235170} {\bibfield  {journal} {\bibinfo
   {journal} {Phys. Rev. B}\ }\textbf {\bibinfo {volume} {102}},\ \bibinfo
  {pages} {235170} (\bibinfo {year} {2020}{\natexlab{b}})}\BibitemShut
  {NoStop}%
\bibitem [{\citenamefont {Elman}\ \emph {et~al.}(2021)\citenamefont {Elman},
  \citenamefont {Chapman},\ and\ \citenamefont {Flammia}}]{FFD-chapman-2021}%
  \BibitemOpen
  \bibfield  {author} {\bibinfo {author} {\bibfnamefont {S.~J.}\ \bibnamefont
  {Elman}}, \bibinfo {author} {\bibfnamefont {A.}~\bibnamefont {Chapman}},\
  and\ \bibinfo {author} {\bibfnamefont {S.~T.}\ \bibnamefont {Flammia}},\
  }\bibfield  {title} {\bibinfo {title} {{Free fermions behind the disguise}},\
  }\href {https://doi.org/10.1007/s00220-021-04220-w} {\bibfield  {journal}
  {\bibinfo  {journal} {Commun. Math. Phys.}\ }\textbf {\bibinfo {volume}
  {388}},\ \bibinfo {pages} {969} (\bibinfo {year} {2021})}\BibitemShut
  {NoStop}%
\bibitem [{\citenamefont {Chapman}\ \emph {et~al.}(2023)\citenamefont
  {Chapman}, \citenamefont {Elman},\ and\ \citenamefont
  {Mann}}]{FFD-chapman-unified-2023}%
  \BibitemOpen
  \bibfield  {author} {\bibinfo {author} {\bibfnamefont {A.}~\bibnamefont
  {Chapman}}, \bibinfo {author} {\bibfnamefont {S.~J.}\ \bibnamefont {Elman}},\
  and\ \bibinfo {author} {\bibfnamefont {R.~L.}\ \bibnamefont {Mann}},\
  }\href@noop {} {\bibinfo {title} {{A Unified Graph-Theoretic Framework for
  Free-Fermion Solvability}}} (\bibinfo {year} {2023}),\ \Eprint
  {https://arxiv.org/abs/2305.15625} {arXiv:2305.15625 [quant-ph]} \BibitemShut
  {NoStop}%
\bibitem [{\citenamefont {Alcaraz}\ \emph {et~al.}(2023)\citenamefont
  {Alcaraz}, \citenamefont {Pimenta},\ and\ \citenamefont
  {Sirker}}]{alcaraz-pimenta-sirker-2023}%
  \BibitemOpen
  \bibfield  {author} {\bibinfo {author} {\bibfnamefont {F.~C.}\ \bibnamefont
  {Alcaraz}}, \bibinfo {author} {\bibfnamefont {R.~A.}\ \bibnamefont
  {Pimenta}},\ and\ \bibinfo {author} {\bibfnamefont {J.}~\bibnamefont
  {Sirker}},\ }\bibfield  {title} {\bibinfo {title} {{Ising analogs of quantum
  spin chains with multispin interactions}},\ }\href
  {https://doi.org/10.1103/PhysRevB.107.235136} {\bibfield  {journal} {\bibinfo
   {journal} {Phys. Rev. B}\ }\textbf {\bibinfo {volume} {107}},\ \bibinfo
  {pages} {235136} (\bibinfo {year} {2023})}\BibitemShut {NoStop}%
\bibitem [{\citenamefont {Fendley}\ and\ \citenamefont
  {Pozsgay}(2024)}]{fendley-pozsgay-2024}%
  \BibitemOpen
  \bibfield  {author} {\bibinfo {author} {\bibfnamefont {P.}~\bibnamefont
  {Fendley}}\ and\ \bibinfo {author} {\bibfnamefont {B.}~\bibnamefont
  {Pozsgay}},\ }\bibfield  {title} {\bibinfo {title} {{Free fermions beyond
  Jordan and Wigner}},\ }\href {https://doi.org/10.21468/SciPostPhys.16.4.102}
  {\bibfield  {journal} {\bibinfo  {journal} {SciPost Phys.}\ }\textbf
  {\bibinfo {volume} {16}},\ \bibinfo {pages} {102} (\bibinfo {year}
  {2024})}\BibitemShut {NoStop}%
\bibitem [{\citenamefont {Vona}\ \emph {et~al.}(2025)\citenamefont {Vona},
  \citenamefont {Mesty\'an},\ and\ \citenamefont
  {Pozsgay}}]{Istvan-ffd-prb-2025}%
  \BibitemOpen
  \bibfield  {author} {\bibinfo {author} {\bibfnamefont {I.}~\bibnamefont
  {Vona}}, \bibinfo {author} {\bibfnamefont {M.}~\bibnamefont {Mesty\'an}},\
  and\ \bibinfo {author} {\bibfnamefont {B.}~\bibnamefont {Pozsgay}},\
  }\bibfield  {title} {\bibinfo {title} {{Exact real-time dynamics with free
  fermions in disguise}},\ }\href {https://doi.org/10.1103/PhysRevB.111.144306}
  {\bibfield  {journal} {\bibinfo  {journal} {Phys. Rev. B}\ }\textbf {\bibinfo
  {volume} {111}},\ \bibinfo {pages} {144306} (\bibinfo {year}
  {2025})}\BibitemShut {NoStop}%
\bibitem [{\citenamefont {Mann}\ \emph {et~al.}(2025)\citenamefont {Mann},
  \citenamefont {Elman}, \citenamefont {Wood},\ and\ \citenamefont
  {Chapman}}]{mann-elman-wood-chapman-parafermion-2025}%
  \BibitemOpen
  \bibfield  {author} {\bibinfo {author} {\bibfnamefont {R.~L.}\ \bibnamefont
  {Mann}}, \bibinfo {author} {\bibfnamefont {S.~J.}\ \bibnamefont {Elman}},
  \bibinfo {author} {\bibfnamefont {D.~R.}\ \bibnamefont {Wood}},\ and\
  \bibinfo {author} {\bibfnamefont {A.}~\bibnamefont {Chapman}},\ }\bibfield
  {title} {\bibinfo {title} {{A graph-theoretic framework for free-parafermion
  solvability}},\ }\href {https://doi.org/10.1098/rspa.2024.0671} {\bibfield
  {journal} {\bibinfo  {journal} {Proc. R. Soc. A}\ }\textbf {\bibinfo {volume}
  {481}},\ \bibinfo {pages} {20240671} (\bibinfo {year} {2025})}\BibitemShut
  {NoStop}%
\bibitem [{\citenamefont {Fukai}\ and\ \citenamefont
  {Pozsgay}(2025)}]{fukai-pozsgay-ffd-circuit-2025}%
  \BibitemOpen
  \bibfield  {author} {\bibinfo {author} {\bibfnamefont {K.}~\bibnamefont
  {Fukai}}\ and\ \bibinfo {author} {\bibfnamefont {B.}~\bibnamefont
  {Pozsgay}},\ }\bibfield  {title} {\bibinfo {title} {{Quantum circuits with
  free fermions in disguise}},\ }\href
  {https://doi.org/10.1088/1751-8121/adcd18} {\bibfield  {journal} {\bibinfo
  {journal} {J. Phys. A}\ }\textbf {\bibinfo {volume} {58}},\ \bibinfo {pages}
  {175202} (\bibinfo {year} {2025})}\BibitemShut {NoStop}%
\bibitem [{\citenamefont {Fukai}\ \emph {et~al.}(2025)\citenamefont {Fukai},
  \citenamefont {Vona},\ and\ \citenamefont {Pozsgay}}]{fukai-claw-2025}%
  \BibitemOpen
  \bibfield  {author} {\bibinfo {author} {\bibfnamefont {K.}~\bibnamefont
  {Fukai}}, \bibinfo {author} {\bibfnamefont {I.}~\bibnamefont {Vona}},\ and\
  \bibinfo {author} {\bibfnamefont {B.}~\bibnamefont {Pozsgay}},\ }\href
  {https://arxiv.org/abs/2508.05789} {\bibinfo {title} {{A free fermions in
  disguise model with claws}}} (\bibinfo {year} {2025}),\ \Eprint
  {https://arxiv.org/abs/2508.05789} {arXiv:2508.05789 [cond-mat.stat-mech]}
  \BibitemShut {NoStop}%
\bibitem [{\citenamefont {Alcaraz}(2025)}]{alcaraz-nonhomogeneous-range-2025}%
  \BibitemOpen
  \bibfield  {author} {\bibinfo {author} {\bibfnamefont {F.~C.}\ \bibnamefont
  {Alcaraz}},\ }\href {https://arxiv.org/abs/2512.08011} {\bibinfo {title}
  {{Free fermionic and parafermionic multispin quantum chains with
  non-homogeneous interacting ranges}}} (\bibinfo {year} {2025}),\ \Eprint
  {https://arxiv.org/abs/2512.08011} {arXiv:2512.08011 [cond-mat.stat-mech]}
  \BibitemShut {NoStop}%
\bibitem [{\citenamefont {Vernier}\ and\ \citenamefont
  {Piroli}(2026)}]{Eric-FFD-hilbert-space-2025}%
  \BibitemOpen
  \bibfield  {author} {\bibinfo {author} {\bibfnamefont {E.}~\bibnamefont
  {Vernier}}\ and\ \bibinfo {author} {\bibfnamefont {L.}~\bibnamefont
  {Piroli}},\ }\bibfield  {title} {\bibinfo {title} {{The Hilbert-space
  structure of free fermions in disguise}},\ }\href
  {https://doi.org/10.1088/1742-5468/ae29f5} {\bibfield  {journal} {\bibinfo
  {journal} {J. Stat. Mech.}\ }\textbf {\bibinfo {volume} {2026}},\ \bibinfo
  {pages} {013101} (\bibinfo {year} {2026})}\BibitemShut {NoStop}%
\bibitem [{\citenamefont {Sz\'asz-Schagrin}\ \emph {et~al.}(2026)\citenamefont
  {Sz\'asz-Schagrin}, \citenamefont {Cristani}, \citenamefont {Piroli},\ and\
  \citenamefont {Vernier}}]{David-ffd-circ-2025}%
  \BibitemOpen
  \bibfield  {author} {\bibinfo {author} {\bibfnamefont {D.}~\bibnamefont
  {Sz\'asz-Schagrin}}, \bibinfo {author} {\bibfnamefont {D.}~\bibnamefont
  {Cristani}}, \bibinfo {author} {\bibfnamefont {L.}~\bibnamefont {Piroli}},\
  and\ \bibinfo {author} {\bibfnamefont {E.}~\bibnamefont {Vernier}},\
  }\bibfield  {title} {\bibinfo {title} {{Construction and simulability of
  quantum circuits with free fermions in disguise}},\ }\href
  {https://doi.org/10.1088/2058-9565/ae390d} {\bibfield  {journal} {\bibinfo
  {journal} {Quantum Sci. Technol.}\ }\textbf {\bibinfo {volume} {11}},\
  \bibinfo {pages} {015044} (\bibinfo {year} {2026})}\BibitemShut {NoStop}%
\bibitem [{\citenamefont {Fukai}\ \emph
  {et~al.}(2026{\natexlab{a}})\citenamefont {Fukai}, \citenamefont {Pozsgay},\
  and\ \citenamefont {Vona}}]{fukai-vona-pozsgay-ffd-eigenstates-paper1-2026}%
  \BibitemOpen
  \bibfield  {author} {\bibinfo {author} {\bibfnamefont {K.}~\bibnamefont
  {Fukai}}, \bibinfo {author} {\bibfnamefont {B.}~\bibnamefont {Pozsgay}},\
  and\ \bibinfo {author} {\bibfnamefont {I.}~\bibnamefont {Vona}},\ }\href
  {https://arxiv.org/abs/2602.03431} {\bibinfo {title} {{Solving models with
  generalized free fermions I: Algebras and eigenstates}}} (\bibinfo {year}
  {2026}{\natexlab{a}}),\ \Eprint {https://arxiv.org/abs/2602.03431}
  {arXiv:2602.03431 [cond-mat.stat-mech]} \BibitemShut {NoStop}%
\bibitem [{\citenamefont {Ruh}\ and\ \citenamefont
  {Elman}(2026)}]{Ruh-Elman-2026-twin-collapse}%
  \BibitemOpen
  \bibfield  {author} {\bibinfo {author} {\bibfnamefont {J.}~\bibnamefont
  {Ruh}}\ and\ \bibinfo {author} {\bibfnamefont {S.~J.}\ \bibnamefont
  {Elman}},\ }\bibfield  {title} {\bibinfo {title} {{Expanding the Class of
  Free Fermions via Twin-Collapse Methods}},\ }\href
  {https://doi.org/10.1103/svfv-vdh5} {\bibfield  {journal} {\bibinfo
  {journal} {Phys. Rev. A}\ }\textbf {\bibinfo {volume} {113}},\ \bibinfo
  {pages} {022456} (\bibinfo {year} {2026})},\ \Eprint
  {https://arxiv.org/abs/2509.09092} {arXiv:2509.09092 [quant-ph]} \BibitemShut
  {NoStop}%
\bibitem [{\citenamefont {Lieb}\ \emph {et~al.}(1961)\citenamefont {Lieb},
  \citenamefont {Schultz},\ and\ \citenamefont {Mattis}}]{LSM}%
  \BibitemOpen
  \bibfield  {author} {\bibinfo {author} {\bibfnamefont {E.}~\bibnamefont
  {Lieb}}, \bibinfo {author} {\bibfnamefont {T.}~\bibnamefont {Schultz}},\ and\
  \bibinfo {author} {\bibfnamefont {D.}~\bibnamefont {Mattis}},\ }\bibfield
  {title} {\bibinfo {title} {{Two soluble models of an antiferromagnetic
  chain}},\ }\href {https://doi.org/10.1016/0003-4916(61)90115-4} {\bibfield
  {journal} {\bibinfo  {journal} {Ann. Phys.}\ }\textbf {\bibinfo {volume}
  {16}},\ \bibinfo {pages} {407} (\bibinfo {year} {1961})}\BibitemShut
  {NoStop}%
\bibitem [{\citenamefont {Pfeuty}(1970)}]{pfeuty1970}%
  \BibitemOpen
  \bibfield  {author} {\bibinfo {author} {\bibfnamefont {P.}~\bibnamefont
  {Pfeuty}},\ }\bibfield  {title} {\bibinfo {title} {{The one-dimensional Ising
  model with a transverse field}},\ }\href
  {https://doi.org/10.1016/0003-4916(70)90270-8} {\bibfield  {journal}
  {\bibinfo  {journal} {Ann. Phys.}\ }\textbf {\bibinfo {volume} {57}},\
  \bibinfo {pages} {79} (\bibinfo {year} {1970})}\BibitemShut {NoStop}%
\bibitem [{\citenamefont {Kitaev}(2006)}]{kitaev2006anyons}%
  \BibitemOpen
  \bibfield  {author} {\bibinfo {author} {\bibfnamefont {A.}~\bibnamefont
  {Kitaev}},\ }\bibfield  {title} {\bibinfo {title} {{Anyons in an exactly
  solved model and beyond}},\ }\href
  {https://doi.org/10.1016/j.aop.2005.10.005} {\bibfield  {journal} {\bibinfo
  {journal} {Ann. Phys.}\ }\textbf {\bibinfo {volume} {321}},\ \bibinfo {pages}
  {2} (\bibinfo {year} {2006})}\BibitemShut {NoStop}%
\bibitem [{\citenamefont {Ogura}\ \emph {et~al.}(2020)\citenamefont {Ogura},
  \citenamefont {Imamura}, \citenamefont {Kameyama}, \citenamefont {Minami},\
  and\ \citenamefont {Sato}}]{ogura-free-fermion-graph-2020}%
  \BibitemOpen
  \bibfield  {author} {\bibinfo {author} {\bibfnamefont {M.}~\bibnamefont
  {Ogura}}, \bibinfo {author} {\bibfnamefont {Y.}~\bibnamefont {Imamura}},
  \bibinfo {author} {\bibfnamefont {N.}~\bibnamefont {Kameyama}}, \bibinfo
  {author} {\bibfnamefont {K.}~\bibnamefont {Minami}},\ and\ \bibinfo {author}
  {\bibfnamefont {M.}~\bibnamefont {Sato}},\ }\bibfield  {title} {\bibinfo
  {title} {{Geometric criterion for solvability of lattice spin systems}},\
  }\href {https://doi.org/10.1103/PhysRevB.102.245118} {\bibfield  {journal}
  {\bibinfo  {journal} {Phys. Rev. B}\ }\textbf {\bibinfo {volume} {102}},\
  \bibinfo {pages} {245118} (\bibinfo {year} {2020})}\BibitemShut {NoStop}%
\bibitem [{\citenamefont {Chapman}\ and\ \citenamefont
  {Flammia}(2020)}]{chapman2020characterization}%
  \BibitemOpen
  \bibfield  {author} {\bibinfo {author} {\bibfnamefont {A.}~\bibnamefont
  {Chapman}}\ and\ \bibinfo {author} {\bibfnamefont {S.~T.}\ \bibnamefont
  {Flammia}},\ }\bibfield  {title} {\bibinfo {title} {{Characterization of
  solvable spin models via graph invariants}},\ }\href
  {https://doi.org/10.22331/q-2020-06-04-278} {\bibfield  {journal} {\bibinfo
  {journal} {Quantum}\ }\textbf {\bibinfo {volume} {4}},\ \bibinfo {pages}
  {278} (\bibinfo {year} {2020})}\BibitemShut {NoStop}%
\bibitem [{\citenamefont {Lindblad}(1976)}]{lindblad-eredeti}%
  \BibitemOpen
  \bibfield  {author} {\bibinfo {author} {\bibfnamefont {G.}~\bibnamefont
  {Lindblad}},\ }\bibfield  {title} {\bibinfo {title} {{On the generators of
  quantum dynamical semigroups}},\ }\href {https://doi.org/10.1007/BF01608499}
  {\bibfield  {journal} {\bibinfo  {journal} {Commun. Math. Phys.}\ }\textbf
  {\bibinfo {volume} {48}},\ \bibinfo {pages} {119} (\bibinfo {year}
  {1976})}\BibitemShut {NoStop}%
\bibitem [{\citenamefont {Rivas}\ \emph {et~al.}(2010)\citenamefont {Rivas},
  \citenamefont {Plato}, \citenamefont {Huelga},\ and\ \citenamefont
  {Plenio}}]{rivas2010markovian}%
  \BibitemOpen
  \bibfield  {author} {\bibinfo {author} {\bibfnamefont {A.}~\bibnamefont
  {Rivas}}, \bibinfo {author} {\bibfnamefont {A.~D.~K.}\ \bibnamefont {Plato}},
  \bibinfo {author} {\bibfnamefont {S.~F.}\ \bibnamefont {Huelga}},\ and\
  \bibinfo {author} {\bibfnamefont {M.~B.}\ \bibnamefont {Plenio}},\ }\bibfield
   {title} {\bibinfo {title} {{Markovian master equations: a critical study}},\
  }\href {https://doi.org/10.1088/1367-2630/12/11/113032} {\bibfield  {journal}
  {\bibinfo  {journal} {New J. Phys.}\ }\textbf {\bibinfo {volume} {12}},\
  \bibinfo {pages} {113032} (\bibinfo {year} {2010})}\BibitemShut {NoStop}%
\bibitem [{\citenamefont {Manzano}(2020)}]{lindblad-intro}%
  \BibitemOpen
  \bibfield  {author} {\bibinfo {author} {\bibfnamefont {D.}~\bibnamefont
  {Manzano}},\ }\bibfield  {title} {\bibinfo {title} {{A short introduction to
  the Lindblad master equation}},\ }\href {https://doi.org/10.1063/1.5115323}
  {\bibfield  {journal} {\bibinfo  {journal} {AIP Advances}\ }\textbf {\bibinfo
  {volume} {10}},\ \bibinfo {pages} {025106} (\bibinfo {year}
  {2020})}\BibitemShut {NoStop}%
\bibitem [{\citenamefont {Bastianello}\ \emph {et~al.}(2020)\citenamefont
  {Bastianello}, \citenamefont {De~Nardis},\ and\ \citenamefont
  {De~Luca}}]{Lindblad-noise}%
  \BibitemOpen
  \bibfield  {author} {\bibinfo {author} {\bibfnamefont {A.}~\bibnamefont
  {Bastianello}}, \bibinfo {author} {\bibfnamefont {J.}~\bibnamefont
  {De~Nardis}},\ and\ \bibinfo {author} {\bibfnamefont {A.}~\bibnamefont
  {De~Luca}},\ }\bibfield  {title} {\bibinfo {title} {{Generalized
  hydrodynamics with dephasing noise}},\ }\href
  {https://doi.org/10.1103/PhysRevB.102.161110} {\bibfield  {journal} {\bibinfo
   {journal} {Phys. Rev. B}\ }\textbf {\bibinfo {volume} {102}},\ \bibinfo
  {pages} {161110} (\bibinfo {year} {2020})}\BibitemShut {NoStop}%
\bibitem [{\citenamefont {Friedman}\ \emph {et~al.}(2020)\citenamefont
  {Friedman}, \citenamefont {Gopalakrishnan},\ and\ \citenamefont
  {Vasseur}}]{vasseur-breaking}%
  \BibitemOpen
  \bibfield  {author} {\bibinfo {author} {\bibfnamefont {A.~J.}\ \bibnamefont
  {Friedman}}, \bibinfo {author} {\bibfnamefont {S.}~\bibnamefont
  {Gopalakrishnan}},\ and\ \bibinfo {author} {\bibfnamefont {R.}~\bibnamefont
  {Vasseur}},\ }\bibfield  {title} {\bibinfo {title} {{Diffusive hydrodynamics
  from integrability breaking}},\ }\href
  {https://doi.org/10.1103/PhysRevB.101.180302} {\bibfield  {journal} {\bibinfo
   {journal} {Phys. Rev. B}\ }\textbf {\bibinfo {volume} {101}},\ \bibinfo
  {pages} {180302} (\bibinfo {year} {2020})}\BibitemShut {NoStop}%
\bibitem [{\citenamefont {Lange}\ \emph {et~al.}(2018)\citenamefont {Lange},
  \citenamefont {Lenar{\v{c}}i{\v{c}}},\ and\ \citenamefont
  {Rosch}}]{zala-tGGE}%
  \BibitemOpen
  \bibfield  {author} {\bibinfo {author} {\bibfnamefont {F.}~\bibnamefont
  {Lange}}, \bibinfo {author} {\bibfnamefont {Z.}~\bibnamefont
  {Lenar{\v{c}}i{\v{c}}}},\ and\ \bibinfo {author} {\bibfnamefont
  {A.}~\bibnamefont {Rosch}},\ }\bibfield  {title} {\bibinfo {title}
  {{Time-dependent generalized Gibbs ensembles in open quantum systems}},\
  }\href {https://doi.org/10.1103/PhysRevB.97.165138} {\bibfield  {journal}
  {\bibinfo  {journal} {Phys. Rev. B}\ }\textbf {\bibinfo {volume} {97}},\
  \bibinfo {pages} {165138} (\bibinfo {year} {2018})}\BibitemShut {NoStop}%
\bibitem [{\citenamefont {Ilievski}(2014)}]{enej-thesis}%
  \BibitemOpen
  \bibfield  {author} {\bibinfo {author} {\bibfnamefont {E.}~\bibnamefont
  {Ilievski}},\ }\href@noop {} {\bibinfo {title} {{Exact solutions of open
  integrable quantum spin chains}}} (\bibinfo {year} {2014}),\ \Eprint
  {https://arxiv.org/abs/1410.1446} {arXiv:1410.1446 [quant-ph]} \BibitemShut
  {NoStop}%
\bibitem [{\citenamefont {Haga}\ \emph {et~al.}(2023)\citenamefont {Haga},
  \citenamefont {Nakagawa}, \citenamefont {Hamazaki},\ and\ \citenamefont
  {Ueda}}]{haga-nakagawa-hamazaki-ueda-incoherentons-2023}%
  \BibitemOpen
  \bibfield  {author} {\bibinfo {author} {\bibfnamefont {T.}~\bibnamefont
  {Haga}}, \bibinfo {author} {\bibfnamefont {M.}~\bibnamefont {Nakagawa}},
  \bibinfo {author} {\bibfnamefont {R.}~\bibnamefont {Hamazaki}},\ and\
  \bibinfo {author} {\bibfnamefont {M.}~\bibnamefont {Ueda}},\ }\bibfield
  {title} {\bibinfo {title} {Quasiparticles of decoherence processes in open
  quantum many-body systems: Incoherentons},\ }\href
  {https://doi.org/10.1103/PhysRevResearch.5.043225} {\bibfield  {journal}
  {\bibinfo  {journal} {Phys. Rev. Res.}\ }\textbf {\bibinfo {volume} {5}},\
  \bibinfo {pages} {043225} (\bibinfo {year} {2023})}\BibitemShut {NoStop}%
\bibitem [{\citenamefont {Rowlands}\ and\ \citenamefont
  {Lamacraft}(2018)}]{rowlands-lamacraft-noisy-spins}%
  \BibitemOpen
  \bibfield  {author} {\bibinfo {author} {\bibfnamefont {D.~A.}\ \bibnamefont
  {Rowlands}}\ and\ \bibinfo {author} {\bibfnamefont {A.}~\bibnamefont
  {Lamacraft}},\ }\bibfield  {title} {\bibinfo {title} {{Noisy Spins and the
  Richardson-Gaudin Model}},\ }\href
  {https://doi.org/10.1103/PhysRevLett.120.090401} {\bibfield  {journal}
  {\bibinfo  {journal} {Phys. Rev. Lett.}\ }\textbf {\bibinfo {volume} {120}},\
  \bibinfo {pages} {090401} (\bibinfo {year} {2018})}\BibitemShut {NoStop}%
\bibitem [{\citenamefont {Essler}\ and\ \citenamefont
  {Piroli}(2020)}]{essler-piroli-lindblad}%
  \BibitemOpen
  \bibfield  {author} {\bibinfo {author} {\bibfnamefont {F.~H.~L.}\
  \bibnamefont {Essler}}\ and\ \bibinfo {author} {\bibfnamefont
  {L.}~\bibnamefont {Piroli}},\ }\bibfield  {title} {\bibinfo {title}
  {{Integrability of one-dimensional Lindbladians from operator-space
  fragmentation}},\ }\href {https://doi.org/10.1103/PhysRevE.102.062210}
  {\bibfield  {journal} {\bibinfo  {journal} {Phys. Rev. E}\ }\textbf {\bibinfo
  {volume} {102}},\ \bibinfo {pages} {062210} (\bibinfo {year}
  {2020})}\BibitemShut {NoStop}%
\bibitem [{\citenamefont {Prosen}(2008)}]{third-quantization}%
  \BibitemOpen
  \bibfield  {author} {\bibinfo {author} {\bibfnamefont {T.}~\bibnamefont
  {Prosen}},\ }\bibfield  {title} {\bibinfo {title} {{Third quantization: a
  general method to solve master equations for quadratic open Fermi systems}},\
  }\href {https://doi.org/10.1088/1367-2630/10/4/043026} {\bibfield  {journal}
  {\bibinfo  {journal} {New J. Phys.}\ }\textbf {\bibinfo {volume} {10}},\
  \bibinfo {pages} {043026} (\bibinfo {year} {2008})}\BibitemShut {NoStop}%
\bibitem [{\citenamefont {Prosen}\ and\ \citenamefont
  {\v{Z}unkovi\v{c}}(2010)}]{third-quantization-2}%
  \BibitemOpen
  \bibfield  {author} {\bibinfo {author} {\bibfnamefont {T.}~\bibnamefont
  {Prosen}}\ and\ \bibinfo {author} {\bibfnamefont {B.}~\bibnamefont
  {\v{Z}unkovi\v{c}}},\ }\bibfield  {title} {\bibinfo {title} {{Exact solution
  of Markovian master equations for quadratic Fermi systems: thermal baths,
  open XY spin chains and non-equilibrium phase transition}},\ }\href
  {https://doi.org/10.1088/1367-2630/12/2/025016} {\bibfield  {journal}
  {\bibinfo  {journal} {New J. Phys.}\ }\textbf {\bibinfo {volume} {12}},\
  \bibinfo {pages} {025016} (\bibinfo {year} {2010})}\BibitemShut {NoStop}%
\bibitem [{\citenamefont {Shibata}\ and\ \citenamefont
  {Katsura}(2019{\natexlab{a}})}]{katsura-lindblad-1}%
  \BibitemOpen
  \bibfield  {author} {\bibinfo {author} {\bibfnamefont {N.}~\bibnamefont
  {Shibata}}\ and\ \bibinfo {author} {\bibfnamefont {H.}~\bibnamefont
  {Katsura}},\ }\bibfield  {title} {\bibinfo {title} {{Dissipative spin chain
  as a non-Hermitian Kitaev ladder}},\ }\href
  {https://doi.org/10.1103/PhysRevB.99.174303} {\bibfield  {journal} {\bibinfo
  {journal} {Phys. Rev. B}\ }\textbf {\bibinfo {volume} {99}},\ \bibinfo
  {pages} {174303} (\bibinfo {year} {2019}{\natexlab{a}})}\BibitemShut
  {NoStop}%
\bibitem [{\citenamefont {Vernier}(2020)}]{eric-lindblad}%
  \BibitemOpen
  \bibfield  {author} {\bibinfo {author} {\bibfnamefont {E.}~\bibnamefont
  {Vernier}},\ }\bibfield  {title} {\bibinfo {title} {{Mixing times and cutoffs
  in open quadratic fermionic systems}},\ }\href
  {https://doi.org/10.21468/SciPostPhys.9.4.049} {\bibfield  {journal}
  {\bibinfo  {journal} {SciPost Phys.}\ }\textbf {\bibinfo {volume} {9}},\
  \bibinfo {pages} {049} (\bibinfo {year} {2020})}\BibitemShut {NoStop}%
\bibitem [{\citenamefont
  {Prosen}(2011{\natexlab{a}})}]{prosen-boundary-lindblad-1}%
  \BibitemOpen
  \bibfield  {author} {\bibinfo {author} {\bibfnamefont {T.}~\bibnamefont
  {Prosen}},\ }\bibfield  {title} {\bibinfo {title} {{Open XXZ Spin Chain:
  Nonequilibrium Steady State and a Strict Bound on Ballistic Transport}},\
  }\href {https://doi.org/10.1103/PhysRevLett.106.217206} {\bibfield  {journal}
  {\bibinfo  {journal} {Phys. Rev. Lett.}\ }\textbf {\bibinfo {volume} {106}},\
  \bibinfo {pages} {217206} (\bibinfo {year} {2011}{\natexlab{a}})}\BibitemShut
  {NoStop}%
\bibitem [{\citenamefont
  {Prosen}(2011{\natexlab{b}})}]{prosen-boundary-lindblad-2}%
  \BibitemOpen
  \bibfield  {author} {\bibinfo {author} {\bibfnamefont {T.}~\bibnamefont
  {Prosen}},\ }\bibfield  {title} {\bibinfo {title} {{Exact Nonequilibrium
  Steady State of a Strongly Driven Open $XXZ$ Chain}},\ }\href
  {https://doi.org/10.1103/PhysRevLett.107.137201} {\bibfield  {journal}
  {\bibinfo  {journal} {Phys. Rev. Lett.}\ }\textbf {\bibinfo {volume} {107}},\
  \bibinfo {pages} {137201} (\bibinfo {year} {2011}{\natexlab{b}})}\BibitemShut
  {NoStop}%
\bibitem [{\citenamefont {Karevski}\ \emph {et~al.}(2013)\citenamefont
  {Karevski}, \citenamefont {Popkov},\ and\ \citenamefont
  {Sch\"utz}}]{boundary-lindblad-mps}%
  \BibitemOpen
  \bibfield  {author} {\bibinfo {author} {\bibfnamefont {D.}~\bibnamefont
  {Karevski}}, \bibinfo {author} {\bibfnamefont {V.}~\bibnamefont {Popkov}},\
  and\ \bibinfo {author} {\bibfnamefont {G.~M.}\ \bibnamefont {Sch\"utz}},\
  }\bibfield  {title} {\bibinfo {title} {{Exact Matrix Product Solution for the
  Boundary-Driven Lindblad XXZ Chain}},\ }\href
  {https://doi.org/10.1103/PhysRevLett.110.047201} {\bibfield  {journal}
  {\bibinfo  {journal} {Phys. Rev. Lett.}\ }\textbf {\bibinfo {volume} {110}},\
  \bibinfo {pages} {047201} (\bibinfo {year} {2013})}\BibitemShut {NoStop}%
\bibitem [{\citenamefont {Prosen}\ \emph {et~al.}(2013)\citenamefont {Prosen},
  \citenamefont {Ilievski},\ and\ \citenamefont
  {Popkov}}]{prosen-exterior-lindblad}%
  \BibitemOpen
  \bibfield  {author} {\bibinfo {author} {\bibfnamefont {T.}~\bibnamefont
  {Prosen}}, \bibinfo {author} {\bibfnamefont {E.}~\bibnamefont {Ilievski}},\
  and\ \bibinfo {author} {\bibfnamefont {V.}~\bibnamefont {Popkov}},\
  }\bibfield  {title} {\bibinfo {title} {{Exterior integrability: Yang-Baxter
  form of non-equilibrium steady-state density operator}},\ }\href
  {https://doi.org/10.1088/1367-2630/15/7/073051} {\bibfield  {journal}
  {\bibinfo  {journal} {New J. Phys.}\ }\textbf {\bibinfo {volume} {15}},\
  \bibinfo {pages} {073051} (\bibinfo {year} {2013})}\BibitemShut {NoStop}%
\bibitem [{\citenamefont {Matsui}\ and\ \citenamefont
  {Prosen}(2017)}]{matsui-prosen-boundary-ness-2017}%
  \BibitemOpen
  \bibfield  {author} {\bibinfo {author} {\bibfnamefont {C.}~\bibnamefont
  {Matsui}}\ and\ \bibinfo {author} {\bibfnamefont {T.}~\bibnamefont
  {Prosen}},\ }\bibfield  {title} {\bibinfo {title} {{Construction of the
  steady state density matrix and quasilocal charges for the spin-1/2 XXZ chain
  with boundary magnetic fields}},\ }\href
  {https://doi.org/10.1088/1751-8121/aa82db} {\bibfield  {journal} {\bibinfo
  {journal} {J. Phys. A}\ }\textbf {\bibinfo {volume} {50}},\ \bibinfo {pages}
  {385201} (\bibinfo {year} {2017})}\BibitemShut {NoStop}%
\bibitem [{\citenamefont {Popkov}\ and\ \citenamefont
  {Sch\"utz}(2017)}]{Popkov-helix-2017}%
  \BibitemOpen
  \bibfield  {author} {\bibinfo {author} {\bibfnamefont {V.}~\bibnamefont
  {Popkov}}\ and\ \bibinfo {author} {\bibfnamefont {G.~M.}\ \bibnamefont
  {Sch\"utz}},\ }\bibfield  {title} {\bibinfo {title} {{Solution of the
  Lindblad equation for spin helix states}},\ }\href
  {https://doi.org/10.1103/PhysRevE.95.042128} {\bibfield  {journal} {\bibinfo
  {journal} {Phys. Rev. E}\ }\textbf {\bibinfo {volume} {95}},\ \bibinfo
  {pages} {042128} (\bibinfo {year} {2017})}\BibitemShut {NoStop}%
\bibitem [{\citenamefont {Matsui}\ and\ \citenamefont
  {Tsuji}(2024)}]{matsui-tsuji-impurity-ness-2024}%
  \BibitemOpen
  \bibfield  {author} {\bibinfo {author} {\bibfnamefont {C.}~\bibnamefont
  {Matsui}}\ and\ \bibinfo {author} {\bibfnamefont {N.}~\bibnamefont {Tsuji}},\
  }\bibfield  {title} {\bibinfo {title} {{Exact steady states of the
  impurity-doped XXZ spin chain coupled to dissipators}},\ }\href
  {https://doi.org/10.1088/1742-5468/ad2b5c} {\bibfield  {journal} {\bibinfo
  {journal} {J. Stat. Mech.}\ }\textbf {\bibinfo {volume} {2024}},\ \bibinfo
  {pages} {033105} (\bibinfo {year} {2024})}\BibitemShut {NoStop}%
\bibitem [{\citenamefont {Bu\v{c}a}\ \emph {et~al.}(2020)\citenamefont
  {Bu\v{c}a}, \citenamefont {Booker}, \citenamefont {Medenjak},\ and\
  \citenamefont {Jaksch}}]{marko-lindblad-1}%
  \BibitemOpen
  \bibfield  {author} {\bibinfo {author} {\bibfnamefont {B.}~\bibnamefont
  {Bu\v{c}a}}, \bibinfo {author} {\bibfnamefont {C.}~\bibnamefont {Booker}},
  \bibinfo {author} {\bibfnamefont {M.}~\bibnamefont {Medenjak}},\ and\
  \bibinfo {author} {\bibfnamefont {D.}~\bibnamefont {Jaksch}},\ }\bibfield
  {title} {\bibinfo {title} {{Bethe ansatz approach for dissipation: exact
  solutions of quantum many-body dynamics under loss}},\ }\href
  {https://doi.org/10.1088/1367-2630/abd124} {\bibfield  {journal} {\bibinfo
  {journal} {New J. Phys.}\ }\textbf {\bibinfo {volume} {22}},\ \bibinfo
  {pages} {123040} (\bibinfo {year} {2020})}\BibitemShut {NoStop}%
\bibitem [{\citenamefont {Nakagawa}\ \emph {et~al.}(2021)\citenamefont
  {Nakagawa}, \citenamefont {Kawakami},\ and\ \citenamefont
  {Ueda}}]{japanok-lindblad}%
  \BibitemOpen
  \bibfield  {author} {\bibinfo {author} {\bibfnamefont {M.}~\bibnamefont
  {Nakagawa}}, \bibinfo {author} {\bibfnamefont {N.}~\bibnamefont {Kawakami}},\
  and\ \bibinfo {author} {\bibfnamefont {M.}~\bibnamefont {Ueda}},\ }\bibfield
  {title} {\bibinfo {title} {{Exact Liouvillian Spectrum of a One-Dimensional
  Dissipative Hubbard Model}},\ }\href
  {https://doi.org/10.1103/PhysRevLett.126.110404} {\bibfield  {journal}
  {\bibinfo  {journal} {Phys. Rev. Lett.}\ }\textbf {\bibinfo {volume} {126}},\
  \bibinfo {pages} {110404} (\bibinfo {year} {2021})}\BibitemShut {NoStop}%
\bibitem [{\citenamefont {Medvedyeva}\ \emph {et~al.}(2016)\citenamefont
  {Medvedyeva}, \citenamefont {Essler},\ and\ \citenamefont
  {Prosen}}]{essler-prosen-lindblad}%
  \BibitemOpen
  \bibfield  {author} {\bibinfo {author} {\bibfnamefont {M.~V.}\ \bibnamefont
  {Medvedyeva}}, \bibinfo {author} {\bibfnamefont {F.~H.~L.}\ \bibnamefont
  {Essler}},\ and\ \bibinfo {author} {\bibfnamefont {T.}~\bibnamefont
  {Prosen}},\ }\bibfield  {title} {\bibinfo {title} {{Exact Bethe Ansatz
  Spectrum of a Tight-Binding Chain with Dephasing Noise}},\ }\href
  {https://doi.org/10.1103/PhysRevLett.117.137202} {\bibfield  {journal}
  {\bibinfo  {journal} {Phys. Rev. Lett.}\ }\textbf {\bibinfo {volume} {117}},\
  \bibinfo {pages} {137202} (\bibinfo {year} {2016})}\BibitemShut {NoStop}%
\bibitem [{\citenamefont {Shibata}\ and\ \citenamefont
  {Katsura}(2019{\natexlab{b}})}]{katsura-lindblad-2}%
  \BibitemOpen
  \bibfield  {author} {\bibinfo {author} {\bibfnamefont {N.}~\bibnamefont
  {Shibata}}\ and\ \bibinfo {author} {\bibfnamefont {H.}~\bibnamefont
  {Katsura}},\ }\bibfield  {title} {\bibinfo {title} {{Dissipative quantum
  Ising chain as a non-Hermitian Ashkin-Teller model}},\ }\href
  {https://doi.org/10.1103/PhysRevB.99.224432} {\bibfield  {journal} {\bibinfo
  {journal} {Phys. Rev. B}\ }\textbf {\bibinfo {volume} {99}},\ \bibinfo
  {pages} {224432} (\bibinfo {year} {2019}{\natexlab{b}})}\BibitemShut
  {NoStop}%
\bibitem [{\citenamefont {Ziolkowska}\ and\ \citenamefont
  {Essler}(2020)}]{essler-lindblad-review}%
  \BibitemOpen
  \bibfield  {author} {\bibinfo {author} {\bibfnamefont {A.~A.}\ \bibnamefont
  {Ziolkowska}}\ and\ \bibinfo {author} {\bibfnamefont {F.~H.~L.}\ \bibnamefont
  {Essler}},\ }\bibfield  {title} {\bibinfo {title} {{Yang-Baxter integrable
  Lindblad equations}},\ }\href {https://doi.org/10.21468/SciPostPhys.8.3.044}
  {\bibfield  {journal} {\bibinfo  {journal} {SciPost Phys.}\ }\textbf
  {\bibinfo {volume} {8}},\ \bibinfo {pages} {044} (\bibinfo {year}
  {2020})}\BibitemShut {NoStop}%
\bibitem [{\citenamefont {de~Leeuw}\ \emph {et~al.}(2021)\citenamefont
  {de~Leeuw}, \citenamefont {Paletta},\ and\ \citenamefont
  {Pozsgay}}]{Leeuw-PRL-2021}%
  \BibitemOpen
  \bibfield  {author} {\bibinfo {author} {\bibfnamefont {M.}~\bibnamefont
  {de~Leeuw}}, \bibinfo {author} {\bibfnamefont {C.}~\bibnamefont {Paletta}},\
  and\ \bibinfo {author} {\bibfnamefont {B.}~\bibnamefont {Pozsgay}},\
  }\bibfield  {title} {\bibinfo {title} {{Constructing Integrable Lindblad
  Superoperators}},\ }\href {https://doi.org/10.1103/PhysRevLett.126.240403}
  {\bibfield  {journal} {\bibinfo  {journal} {Phys. Rev. Lett.}\ }\textbf
  {\bibinfo {volume} {126}},\ \bibinfo {pages} {240403} (\bibinfo {year}
  {2021})}\BibitemShut {NoStop}%
\bibitem [{\citenamefont {Ekman}\ and\ \citenamefont
  {Bergholtz}(2024)}]{PhysRevResearch.6.L032067}%
  \BibitemOpen
  \bibfield  {author} {\bibinfo {author} {\bibfnamefont {C.}~\bibnamefont
  {Ekman}}\ and\ \bibinfo {author} {\bibfnamefont {E.~J.}\ \bibnamefont
  {Bergholtz}},\ }\bibfield  {title} {\bibinfo {title} {{Liouvillian skin
  effects and fragmented condensates in an integrable dissipative Bose-Hubbard
  model}},\ }\href {https://doi.org/10.1103/PhysRevResearch.6.L032067}
  {\bibfield  {journal} {\bibinfo  {journal} {Phys. Rev. Res.}\ }\textbf
  {\bibinfo {volume} {6}},\ \bibinfo {pages} {L032067} (\bibinfo {year}
  {2024})}\BibitemShut {NoStop}%
\bibitem [{\citenamefont {\ifmmode \check{Z}\else
  \v{Z}\fi{}nidari\ifmmode~\check{c}\else
  \v{c}\fi{}}(2014)}]{znidaric-pre-2014}%
  \BibitemOpen
  \bibfield  {author} {\bibinfo {author} {\bibfnamefont {M.}~\bibnamefont
  {\ifmmode \check{Z}\else \v{Z}\fi{}nidari\ifmmode~\check{c}\else
  \v{c}\fi{}}},\ }\bibfield  {title} {\bibinfo {title} {{Large-deviation
  statistics of a diffusive quantum spin chain and the additivity principle}},\
  }\href {https://doi.org/10.1103/PhysRevE.89.042140} {\bibfield  {journal}
  {\bibinfo  {journal} {Phys. Rev. E}\ }\textbf {\bibinfo {volume} {89}},\
  \bibinfo {pages} {042140} (\bibinfo {year} {2014})}\BibitemShut {NoStop}%
\bibitem [{\citenamefont {\ifmmode \check{Z}\else
  \v{Z}\fi{}nidari\ifmmode~\check{c}\else
  \v{c}\fi{}}(2015)}]{znidaric-pre-2015}%
  \BibitemOpen
  \bibfield  {author} {\bibinfo {author} {\bibfnamefont {M.}~\bibnamefont
  {\ifmmode \check{Z}\else \v{Z}\fi{}nidari\ifmmode~\check{c}\else
  \v{c}\fi{}}},\ }\bibfield  {title} {\bibinfo {title} {{Relaxation times of
  dissipative many-body quantum systems}},\ }\href
  {https://doi.org/10.1103/PhysRevE.92.042143} {\bibfield  {journal} {\bibinfo
  {journal} {Phys. Rev. E}\ }\textbf {\bibinfo {volume} {92}},\ \bibinfo
  {pages} {042143} (\bibinfo {year} {2015})}\BibitemShut {NoStop}%
\bibitem [{Note1()}]{Note1}%
  \BibitemOpen
  \bibinfo {note} {Fendley's original model is defined on an $(M+2)$ site spin
  chain with local terms $h_j=\sigma _j^z \sigma _{j+1}^z \sigma _{j+2}^x$ for
  $j=1,\protect \ldots ,M$, so that $H=\DOTSB \sum@ \slimits@ _{j=1}^{M} b_j
  h_j$. The difference amounts only to a trivial degeneracy associated with the
  first two spins and does not affect the essential structure of the
  model.}\BibitemShut {Stop}%
\bibitem [{\citenamefont {Jordan}\ and\ \citenamefont
  {Wigner}(1993)}]{jordan1993}%
  \BibitemOpen
  \bibfield  {author} {\bibinfo {author} {\bibfnamefont {P.}~\bibnamefont
  {Jordan}}\ and\ \bibinfo {author} {\bibfnamefont {E.~P.}\ \bibnamefont
  {Wigner}},\ }\bibinfo {title} {{{\"U}ber das Paulische
  {\"A}quivalenzverbot}},\ in\ \href
  {https://doi.org/10.1007/978-3-662-02781-3_9} {\emph {\bibinfo {booktitle}
  {The Collected Works of Eugene Paul Wigner: Part A: The Scientific
  Papers}}},\ \bibinfo {editor} {edited by\ \bibinfo {editor} {\bibfnamefont
  {A.~S.}\ \bibnamefont {Wightman}}}\ (\bibinfo  {publisher} {Springer Berlin
  Heidelberg},\ \bibinfo {address} {Berlin, Heidelberg},\ \bibinfo {year}
  {1993})\ pp.\ \bibinfo {pages} {109--129}\BibitemShut {NoStop}%
\bibitem [{Note2()}]{Note2}%
  \BibitemOpen
  \bibinfo {note} {The corresponding real-rootedness result was first proved
  for the unweighted independence polynomial~\cite {chudnovsky-seymour}; see
  also Ref.~\cite {bencs-independence-polynomial}. Refs.~\cite
  {engstrom-weighted,leake-ryder-independence} extend this real-rootedness
  result to weighted independence polynomials. Refs.~\cite
  {engstrom-weighted,leake-ryder-independence} define the independence
  polynomial as $\DOTSB \sum@ \slimits@ _{S} x^{|S|} \DOTSB \prod@ \slimits@
  b_{\protect \bm {j}}^2$ without the $(-1)^{|S|}$ sign, and accordingly state
  that the roots are real \protect \emph {negative}. Our convention with
  $(-x)^{|S|}$ absorbs this sign, making the roots positive.}\BibitemShut
  {Stop}%
\bibitem [{\citenamefont {Chudnovsky}\ \emph {et~al.}(2021)\citenamefont
  {Chudnovsky}, \citenamefont {Scott}, \citenamefont {Seymour},\ and\
  \citenamefont {Spirkl}}]{evenhole-simplicial}%
  \BibitemOpen
  \bibfield  {author} {\bibinfo {author} {\bibfnamefont {M.}~\bibnamefont
  {Chudnovsky}}, \bibinfo {author} {\bibfnamefont {A.}~\bibnamefont {Scott}},
  \bibinfo {author} {\bibfnamefont {P.}~\bibnamefont {Seymour}},\ and\ \bibinfo
  {author} {\bibfnamefont {S.}~\bibnamefont {Spirkl}},\ }\bibfield  {title}
  {\bibinfo {title} {{A note on simplicial cliques}},\ }\href
  {https://doi.org/10.1016/j.disc.2021.112470} {\bibfield  {journal} {\bibinfo
  {journal} {Discrete Math.}\ }\textbf {\bibinfo {volume} {344}},\ \bibinfo
  {pages} {112470} (\bibinfo {year} {2021})}\BibitemShut {NoStop}%
\bibitem [{\citenamefont {Engstr{\"o}m}(2007)}]{engstrom-weighted}%
  \BibitemOpen
  \bibfield  {author} {\bibinfo {author} {\bibfnamefont {A.}~\bibnamefont
  {Engstr{\"o}m}},\ }\bibfield  {title} {\bibinfo {title} {{Inequalities on
  Well-Distributed Point Sets on Circles}},\ }\href {https://emis.de/ft/14440}
  {\bibfield  {journal} {\bibinfo  {journal} {J. Inequal. Pure Appl. Math.}\
  }\textbf {\bibinfo {volume} {8}},\ \bibinfo {pages} {Art.~34} (\bibinfo
  {year} {2007})}\BibitemShut {NoStop}%
\bibitem [{\citenamefont {Leake}\ and\ \citenamefont
  {Ryder}(2019)}]{leake-ryder-independence}%
  \BibitemOpen
  \bibfield  {author} {\bibinfo {author} {\bibfnamefont {J.~D.}\ \bibnamefont
  {Leake}}\ and\ \bibinfo {author} {\bibfnamefont {N.~R.}\ \bibnamefont
  {Ryder}},\ }\bibfield  {title} {\bibinfo {title} {{Generalizations of the
  Matching Polynomial to the Multivariate Independence Polynomial}},\ }\href
  {https://doi.org/10.5802/alco.63} {\bibfield  {journal} {\bibinfo  {journal}
  {Algebr. Comb.}\ }\textbf {\bibinfo {volume} {2}},\ \bibinfo {pages} {781}
  (\bibinfo {year} {2019})}\BibitemShut {NoStop}%
\bibitem [{\citenamefont {Bilstein}\ and\ \citenamefont
  {Wehefritz}(1999)}]{Ulrich-Bilstein-1999}%
  \BibitemOpen
  \bibfield  {author} {\bibinfo {author} {\bibfnamefont {U.}~\bibnamefont
  {Bilstein}}\ and\ \bibinfo {author} {\bibfnamefont {B.}~\bibnamefont
  {Wehefritz}},\ }\bibfield  {title} {\bibinfo {title} {{The $XX$-model with
  boundaries: Part I. Diagonalization of the finite chain}},\ }\href
  {https://doi.org/10.1088/0305-4470/32/2/001} {\bibfield  {journal} {\bibinfo
  {journal} {J. Phys. A}\ }\textbf {\bibinfo {volume} {32}},\ \bibinfo {pages}
  {191} (\bibinfo {year} {1999})}\BibitemShut {NoStop}%
\bibitem [{Note3()}]{Note3}%
  \BibitemOpen
  \bibinfo {note} {At fine-tuned values of $b_j$, additional accidental
  degeneracies may occur. Throughout this work, ``generic couplings'' means
  that such nongeneric degeneracies are excluded.}\BibitemShut {Stop}%
\bibitem [{\citenamefont {Prosen}\ and\ \citenamefont
  {Pi\v{z}orn}(2008)}]{prosen-pizorn-2008}%
  \BibitemOpen
  \bibfield  {author} {\bibinfo {author} {\bibfnamefont {T.}~\bibnamefont
  {Prosen}}\ and\ \bibinfo {author} {\bibfnamefont {I.}~\bibnamefont
  {Pi\v{z}orn}},\ }\bibfield  {title} {\bibinfo {title} {{Quantum Phase
  Transition in a Far-from-Equilibrium Steady State of an XY Spin Chain}},\
  }\href {https://doi.org/10.1103/PhysRevLett.101.105701} {\bibfield  {journal}
  {\bibinfo  {journal} {Phys. Rev. Lett.}\ }\textbf {\bibinfo {volume} {101}},\
  \bibinfo {pages} {105701} (\bibinfo {year} {2008})}\BibitemShut {NoStop}%
\bibitem [{\citenamefont {Medvedyeva}\ and\ \citenamefont
  {Kehrein}(2014)}]{medvedyeva-kehrein-2014}%
  \BibitemOpen
  \bibfield  {author} {\bibinfo {author} {\bibfnamefont {M.~V.}\ \bibnamefont
  {Medvedyeva}}\ and\ \bibinfo {author} {\bibfnamefont {S.}~\bibnamefont
  {Kehrein}},\ }\bibfield  {title} {\bibinfo {title} {{Power-law approach to
  steady state in open lattices of noninteracting electrons}},\ }\href
  {https://doi.org/10.1103/PhysRevB.90.205410} {\bibfield  {journal} {\bibinfo
  {journal} {Phys. Rev. B}\ }\textbf {\bibinfo {volume} {90}},\ \bibinfo
  {pages} {205410} (\bibinfo {year} {2014})}\BibitemShut {NoStop}%
\bibitem [{\citenamefont {Shibata}\ and\ \citenamefont
  {Katsura}(2020)}]{shibata-boundary-dephasing-2020}%
  \BibitemOpen
  \bibfield  {author} {\bibinfo {author} {\bibfnamefont {N.}~\bibnamefont
  {Shibata}}\ and\ \bibinfo {author} {\bibfnamefont {H.}~\bibnamefont
  {Katsura}},\ }\bibfield  {title} {\bibinfo {title} {{Quantum Ising chain with
  boundary dephasing}},\ }\href {https://doi.org/10.1093/ptep/ptaa131}
  {\bibfield  {journal} {\bibinfo  {journal} {Prog. Theor. Exp. Phys.}\
  }\textbf {\bibinfo {volume} {2020}},\ \bibinfo {pages} {12A108} (\bibinfo
  {year} {2020})}\BibitemShut {NoStop}%
\bibitem [{\citenamefont {Fukai}\ \emph
  {et~al.}(2026{\natexlab{b}})\citenamefont {Fukai}, \citenamefont {Pozsgay},\
  and\ \citenamefont {Vona}}]{fukai-vona-pozsgay-ffd-eigenstates-paper2-2026}%
  \BibitemOpen
  \bibfield  {author} {\bibinfo {author} {\bibfnamefont {K.}~\bibnamefont
  {Fukai}}, \bibinfo {author} {\bibfnamefont {B.}~\bibnamefont {Pozsgay}},\
  and\ \bibinfo {author} {\bibfnamefont {I.}~\bibnamefont {Vona}},\ }\bibfield
  {title} {\bibinfo {title} {{Solving models with generalized free fermions II:
  Path-Product Expansion and Local Conserved Charges}}} (\bibinfo {year}
  {2026}{\natexlab{b}}),\ \bibinfo {note} {manuscript in
  preparation}\BibitemShut {NoStop}%
\bibitem [{\citenamefont {Syassen}\ \emph {et~al.}(2008)\citenamefont
  {Syassen}, \citenamefont {Bauer}, \citenamefont {Lettner}, \citenamefont
  {Volz}, \citenamefont {Dietze}, \citenamefont {{Garc{\'i}a-Ripoll}},
  \citenamefont {Cirac}, \citenamefont {Rempe},\ and\ \citenamefont
  {D{\"u}rr}}]{syassen-StrongDissipationInhibits-2008}%
  \BibitemOpen
  \bibfield  {author} {\bibinfo {author} {\bibfnamefont {N.}~\bibnamefont
  {Syassen}}, \bibinfo {author} {\bibfnamefont {D.~M.}\ \bibnamefont {Bauer}},
  \bibinfo {author} {\bibfnamefont {M.}~\bibnamefont {Lettner}}, \bibinfo
  {author} {\bibfnamefont {T.}~\bibnamefont {Volz}}, \bibinfo {author}
  {\bibfnamefont {D.}~\bibnamefont {Dietze}}, \bibinfo {author} {\bibfnamefont
  {J.~J.}\ \bibnamefont {{Garc{\'i}a-Ripoll}}}, \bibinfo {author}
  {\bibfnamefont {J.~I.}\ \bibnamefont {Cirac}}, \bibinfo {author}
  {\bibfnamefont {G.}~\bibnamefont {Rempe}},\ and\ \bibinfo {author}
  {\bibfnamefont {S.}~\bibnamefont {D{\"u}rr}},\ }\bibfield  {title} {\bibinfo
  {title} {Strong {{Dissipation Inhibits Losses}} and {{Induces Correlations}}
  in {{Cold Molecular Gases}}},\ }\href
  {https://doi.org/10.1126/science.1155309} {\bibfield  {journal} {\bibinfo
  {journal} {Science}\ }\textbf {\bibinfo {volume} {320}},\ \bibinfo {pages}
  {1329} (\bibinfo {year} {2008})}\BibitemShut {NoStop}%
\bibitem [{\citenamefont {Yan}\ \emph {et~al.}(2013)\citenamefont {Yan},
  \citenamefont {Moses}, \citenamefont {Gadway}, \citenamefont {Covey},
  \citenamefont {Hazzard}, \citenamefont {Rey}, \citenamefont {Jin},\ and\
  \citenamefont {Ye}}]{yan-ObservationDipolarSpinexchange-2013}%
  \BibitemOpen
  \bibfield  {author} {\bibinfo {author} {\bibfnamefont {B.}~\bibnamefont
  {Yan}}, \bibinfo {author} {\bibfnamefont {S.~A.}\ \bibnamefont {Moses}},
  \bibinfo {author} {\bibfnamefont {B.}~\bibnamefont {Gadway}}, \bibinfo
  {author} {\bibfnamefont {J.~P.}\ \bibnamefont {Covey}}, \bibinfo {author}
  {\bibfnamefont {K.~R.~A.}\ \bibnamefont {Hazzard}}, \bibinfo {author}
  {\bibfnamefont {A.~M.}\ \bibnamefont {Rey}}, \bibinfo {author} {\bibfnamefont
  {D.~S.}\ \bibnamefont {Jin}},\ and\ \bibinfo {author} {\bibfnamefont
  {J.}~\bibnamefont {Ye}},\ }\bibfield  {title} {\bibinfo {title} {Observation
  of dipolar spin-exchange interactions with lattice-confined polar
  molecules},\ }\href {https://doi.org/10.1038/nature12483} {\bibfield
  {journal} {\bibinfo  {journal} {Nature}\ }\textbf {\bibinfo {volume} {501}},\
  \bibinfo {pages} {521} (\bibinfo {year} {2013})}\BibitemShut {NoStop}%
\bibitem [{Note4()}]{Note4}%
  \BibitemOpen
  \bibinfo {note} {ECF is a special case because every even-hole-free claw-free
  graph has a simplicial clique~\cite {evenhole-simplicial}.}\BibitemShut
  {Stop}%
\bibitem [{\citenamefont {Chudnovsky}\ and\ \citenamefont
  {Seymour}(2007)}]{chudnovsky-seymour}%
  \BibitemOpen
  \bibfield  {author} {\bibinfo {author} {\bibfnamefont {M.}~\bibnamefont
  {Chudnovsky}}\ and\ \bibinfo {author} {\bibfnamefont {P.}~\bibnamefont
  {Seymour}},\ }\bibfield  {title} {\bibinfo {title} {{The roots of the
  independence polynomial of a clawfree graph}},\ }\href
  {https://doi.org/10.1016/j.jctb.2006.06.001} {\bibfield  {journal} {\bibinfo
  {journal} {J. Combin. Theory Ser. B}\ }\textbf {\bibinfo {volume} {97}},\
  \bibinfo {pages} {350} (\bibinfo {year} {2007})}\BibitemShut {NoStop}%
\bibitem [{\citenamefont {Bencs}(2018)}]{bencs-independence-polynomial}%
  \BibitemOpen
  \bibfield  {author} {\bibinfo {author} {\bibfnamefont {F.}~\bibnamefont
  {Bencs}},\ }\bibfield  {title} {\bibinfo {title} {{Christoffel--Darboux type
  identities for the independence polynomial}},\ }\href
  {https://doi.org/10.1017/S0963548318000135} {\bibfield  {journal} {\bibinfo
  {journal} {Combin. Probab. Comput.}\ }\textbf {\bibinfo {volume} {27}},\
  \bibinfo {pages} {716} (\bibinfo {year} {2018})}\BibitemShut {NoStop}%
\bibitem [{\citenamefont {Baumgartner}\ and\ \citenamefont
  {Narnhofer}(2008)}]{baumgartner-narnhofer-2008}%
  \BibitemOpen
  \bibfield  {author} {\bibinfo {author} {\bibfnamefont {B.}~\bibnamefont
  {Baumgartner}}\ and\ \bibinfo {author} {\bibfnamefont {H.}~\bibnamefont
  {Narnhofer}},\ }\bibfield  {title} {\bibinfo {title} {{Analysis of quantum
  semigroups with GKS--Lindblad generators: II. General}},\ }\href
  {https://doi.org/10.1088/1751-8113/41/39/395303} {\bibfield  {journal}
  {\bibinfo  {journal} {J. Phys. A}\ }\textbf {\bibinfo {volume} {41}},\
  \bibinfo {pages} {395303} (\bibinfo {year} {2008})}\BibitemShut {NoStop}%
\bibitem [{\citenamefont {Bu\v{c}a}\ and\ \citenamefont
  {Prosen}(2012)}]{buca-prosen-2012}%
  \BibitemOpen
  \bibfield  {author} {\bibinfo {author} {\bibfnamefont {B.}~\bibnamefont
  {Bu\v{c}a}}\ and\ \bibinfo {author} {\bibfnamefont {T.}~\bibnamefont
  {Prosen}},\ }\bibfield  {title} {\bibinfo {title} {{A note on symmetry
  reductions of the Lindblad equation: transport in constrained open spin
  chains}},\ }\href {https://doi.org/10.1088/1367-2630/14/7/073007} {\bibfield
  {journal} {\bibinfo  {journal} {New J. Phys.}\ }\textbf {\bibinfo {volume}
  {14}},\ \bibinfo {pages} {073007} (\bibinfo {year} {2012})}\BibitemShut
  {NoStop}%
\bibitem [{\citenamefont {Albert}\ and\ \citenamefont
  {Jiang}(2014)}]{albert-jiang-2014}%
  \BibitemOpen
  \bibfield  {author} {\bibinfo {author} {\bibfnamefont {V.~V.}\ \bibnamefont
  {Albert}}\ and\ \bibinfo {author} {\bibfnamefont {L.}~\bibnamefont
  {Jiang}},\ }\bibfield  {title} {\bibinfo {title} {{Symmetries and conserved
  quantities in Lindblad master equations}},\ }\href
  {https://doi.org/10.1103/PhysRevA.89.022118} {\bibfield  {journal} {\bibinfo
  {journal} {Phys. Rev. A}\ }\textbf {\bibinfo {volume} {89}},\ \bibinfo
  {pages} {022118} (\bibinfo {year} {2014})}\BibitemShut {NoStop}%
\bibitem [{\citenamefont {Nigro}(2019)}]{nigro-2019}%
  \BibitemOpen
  \bibfield  {author} {\bibinfo {author} {\bibfnamefont {D.}~\bibnamefont
  {Nigro}},\ }\bibfield  {title} {\bibinfo {title} {{On the uniqueness of the
  steady-state solution of the Lindblad--Gorini--Kossakowski--Sudarshan
  equation}},\ }\href {https://doi.org/10.1088/1742-5468/ab0c1c} {\bibfield
  {journal} {\bibinfo  {journal} {J. Stat. Mech.}\ }\textbf {\bibinfo {volume}
  {2019}},\ \bibinfo {pages} {043202} (\bibinfo {year} {2019})}\BibitemShut
  {NoStop}%
\bibitem [{\citenamefont {Iemini}\ \emph {et~al.}(2018)\citenamefont {Iemini},
  \citenamefont {Russomanno}, \citenamefont {Keeling}, \citenamefont
  {Schir\`o}, \citenamefont {Rossini},\ and\ \citenamefont
  {Fazio}}]{iemini-boundary-time-crystals-2018}%
  \BibitemOpen
  \bibfield  {author} {\bibinfo {author} {\bibfnamefont {F.}~\bibnamefont
  {Iemini}}, \bibinfo {author} {\bibfnamefont {A.}~\bibnamefont {Russomanno}},
  \bibinfo {author} {\bibfnamefont {J.}~\bibnamefont {Keeling}}, \bibinfo
  {author} {\bibfnamefont {M.}~\bibnamefont {Schir\`o}}, \bibinfo {author}
  {\bibfnamefont {D.}~\bibnamefont {Rossini}},\ and\ \bibinfo {author}
  {\bibfnamefont {R.}~\bibnamefont {Fazio}},\ }\bibfield  {title} {\bibinfo
  {title} {{Boundary time crystals}},\ }\href
  {https://doi.org/10.1103/PhysRevLett.121.035301} {\bibfield  {journal}
  {\bibinfo  {journal} {Phys. Rev. Lett.}\ }\textbf {\bibinfo {volume}
  {121}},\ \bibinfo {pages} {035301} (\bibinfo {year} {2018})}\BibitemShut
  {NoStop}%
\bibitem [{\citenamefont {Bu\v{c}a}\ \emph {et~al.}(2019)\citenamefont
  {Bu\v{c}a}, \citenamefont {Tindall},\ and\ \citenamefont
  {Jaksch}}]{buca-tindall-jaksch-2019}%
  \BibitemOpen
  \bibfield  {author} {\bibinfo {author} {\bibfnamefont {B.}~\bibnamefont
  {Bu\v{c}a}}, \bibinfo {author} {\bibfnamefont {J.}~\bibnamefont {Tindall}},\
  and\ \bibinfo {author} {\bibfnamefont {D.}~\bibnamefont {Jaksch}},\ }\bibfield
  {title} {\bibinfo {title} {{Non-stationary coherent quantum many-body
  dynamics through dissipation}},\ }\href
  {https://doi.org/10.1038/s41467-019-09757-y} {\bibfield  {journal} {\bibinfo
  {journal} {Nat. Commun.}\ }\textbf {\bibinfo {volume} {10}},\ \bibinfo
  {pages} {1730} (\bibinfo {year} {2019})}\BibitemShut {NoStop}%
\bibitem [{\citenamefont {Bu\v{c}a}\ \emph {et~al.}(2022)\citenamefont
  {Bu\v{c}a}, \citenamefont {Booker},\ and\ \citenamefont
  {Jaksch}}]{buca-booker-jaksch-2022}%
  \BibitemOpen
  \bibfield  {author} {\bibinfo {author} {\bibfnamefont {B.}~\bibnamefont
  {Bu\v{c}a}}, \bibinfo {author} {\bibfnamefont {C.}~\bibnamefont {Booker}},\
  and\ \bibinfo {author} {\bibfnamefont {D.}~\bibnamefont {Jaksch}},\ }\bibfield
  {title} {\bibinfo {title} {{Algebraic theory of quantum synchronization and
  limit cycles under dissipation}},\ }\href
  {https://doi.org/10.21468/SciPostPhys.12.3.097} {\bibfield  {journal}
  {\bibinfo  {journal} {SciPost Phys.}\ }\textbf {\bibinfo {volume} {12}},\
  \bibinfo {pages} {097} (\bibinfo {year} {2022})}\BibitemShut {NoStop}%
\bibitem [{Note5()}]{Note5}%
  \BibitemOpen
  \bibinfo {note} {We expect this coprimality condition to hold for generic
  couplings. In the earlier FFD literature~\cite
  {FFD-fendley-2019,FFD-chapman-2021,FFD-chapman-unified-2023}, coprimality
  appears to be assumed implicitly; when $P_G$ and $P_{G \setminus K_s}$ share
  a root $x_k=u_k^2$, the normalization factor $\protect \mathcal {N}_k$ in
  Eq.~\protect \eqref {eq:free-fermion-mode} vanishes, but this degenerate case
  does not appear to be addressed. Such a coincidence can be engineered by fine
  tuning: when $G$ is the path graph $1$--$2$--$3$--$4$ with uniform couplings
  and $K_s=\{2,3\}$, $P_G(x)=1-4x+3x^2$ and $P_{G\setminus K_s}(x)=(1-x)^2$
  share the root $x=1$. At a shared root $x_\ast =u_\ast ^2$, $\protect
  \polytilpm {G}{\pm }{\pm u_\ast }=0$ and the single-particle energies
  $\protect \liouvillian {\varepsilon }=\pm 1/u_\ast $ are real, leading to
  nonzero Liouvillian eigenvalues on the imaginary axis. The resulting
  nondecaying oscillatory sector, constrained by Lindbladian symmetries and
  invariant subspaces~\cite
  {baumgartner-narnhofer-2008,buca-prosen-2012,albert-jiang-2014,nigro-2019},
  can produce persistent oscillations in the long-time dynamics~\cite
  {iemini-boundary-time-crystals-2018,buca-tindall-jaksch-2019,buca-booker-jaksch-2022}.}\BibitemShut
  {Stop}%
\end{thebibliography}%

\section*{End Matter}
\appendix

\secc{Defining representation of the graph-Clifford algebra}
Choose $V(G)=\{1,\ldots,M\}$ with $M\equiv |V(G)|$.
Ref.~\cite{fukai-vona-pozsgay-ffd-eigenstates-paper1-2026} gives a canonical operator-state correspondence that realizes the abstract graph-Clifford algebra on a spin-$1/2$ chain of length $M$.
One of the two defining spin-chain representations is
\begin{align}
    h_j
    =
    \qty(\prod_{\ell<j} (\sigma_\ell^z)^{A_{\ell j}}) \sigma_j^x.
    \label{eq:defining-representation}
\end{align}
This can be viewed as a graph-dependent generalization of the Jordan-Wigner transformation.
A direct check shows that Eq.~\eqref{eq:defining-representation} satisfies $(h_j)^2=\mathbb{1}$ and reproduces the commutation rule in Eq.~\eqref{eq:graph-clifford}, so every graph-Clifford algebra has a corresponding spin-chain realization on $|V(G)|$ spins.
If $K_s=\{j_1,\ldots,j_r\}$ is a clique, the associated edge operator is represented by
\begin{align}
    \chi
    =
    \sigma_{j_1}^z \cdots \sigma_{j_r}^z,
    \label{eq:defining-representation-edge}
\end{align}
which anticommutes with $h_j$ for $j \in K_s$ and commutes with the rest.

\secc{Normalization of the hidden fermion modes}
The normalization in Eq.~\eqref{eq:free-fermion-mode} is the graph-theoretic generalization of the formula derived in Refs.~\cite{FFD-fendley-2019,FFD-chapman-unified-2023}:
\begin{align}
    \mathcal{N}_k^2
    & =
    -16u_k^2 P_{G\setminus K_s}(u_k^2) P_G'(u_k^2)\,.
    \label{eq:mode-normalization}
\end{align}
Equation~\eqref{eq:mode-normalization} shows that $\mathcal{N}_k=0$ if $P_G$ and $P_{G\setminus K_s}$ share the root $u_k^2$.
Throughout this work, we exclude this nongeneric case~\footnote{
    We expect this shared-root exclusion to hold for generic couplings.
    In the earlier FFD literature~\cite{FFD-fendley-2019,FFD-chapman-2021,FFD-chapman-unified-2023}, it appears to be assumed implicitly; if $P_G(x)$ and $P_{G\setminus K_s}(x)$ share a root $x_\ast$, the standard mode-normalization factor vanishes.
    Such a coincidence can be engineered by fine tuning: for the path graph $1$--$2$--$3$--$4$ with uniform couplings and $K_s=\{2,3\}$, the two polynomials share the root $x=1$.
    At such a shared root $x_\ast$, $\protect\polytilpm{G}{\pm}{\pm\sqrt{x_\ast}}=0$ and the single-particle energies $\protect\liouvillian{\varepsilon}_k=\pm 1/\sqrt{x_\ast}$ are real, giving nonzero Liouvillian eigenvalues on the imaginary axis and hence a nondecaying oscillatory sector~\cite{baumgartner-narnhofer-2008,buca-prosen-2012,albert-jiang-2014,nigro-2019,iemini-boundary-time-crystals-2018,buca-tindall-jaksch-2019,buca-booker-jaksch-2022}.
}.

For the Liouvillian modes, applying the same normalization formula to the doubled graph $\liouvillian{G}$ and the simplicial clique $\liouvillian{K}_s^{(1)}$, and using $P_{\liouvillian{G}}(u^2)=\polytilpm{G}{+}{u}\polytilpm{G}{-}{u}$ together with $P_{\liouvillian{G}\setminus \liouvillian{K}_s^{(1)}}(u^2)=P_{G\setminus K_s}(u^2)P_G(u^2)$, gives
\begin{align}
    \Ntil_k^2
    & =
    -8\liouvillian{u}_k P_{G\setminus K_s}(\liouvillian{u}_k^2) P_G(\liouvillian{u}_k^2) \polytilpm{G}{-}{\liouvillian{u}_k} \qty[\partial_u \polytilpm{G}{+}{u}]_{u=\liouvillian{u}_k}
    \nonumber \\
    & =
    16\gamma^2 \liouvillian{u}_k^3 P_{G\setminus K_s}(\liouvillian{u}_k^2)^3 \qty[\partial_u \polytilpm{G}{+}{u}]_{u=\liouvillian{u}_k}\,.
    \label{eq:liouvillian-mode-normalization}
\end{align}
To simplify the first line, we use that $\liouvillian{u}_k$ is a root of $\polytilpm{G}{+}{u}=P_G(u^2)+\imi\gamma u P_{G\setminus K_s}(u^2)$, which gives
\begin{align}
    P_G(\liouvillian{u}_k^2)
    & =
    -\imi\gamma \liouvillian{u}_k P_{G\setminus K_s}(\liouvillian{u}_k^2),
    \nonumber \\
    \polytilpm{G}{-}{\liouvillian{u}_k}
    & =
    P_G(\liouvillian{u}_k^2)-\imi\gamma \liouvillian{u}_k P_{G\setminus K_s}(\liouvillian{u}_k^2)
    \nonumber \\
    & =
    -2\imi\gamma \liouvillian{u}_k P_{G\setminus K_s}(\liouvillian{u}_k^2).
    \nonumber
\end{align}

\secc{Derivation of the correlation function}
We derive Eq.~\eqref{eq:Bt-residue} for a general ECF graph $G$.
For the Krylov basis $\phi_{j+1}=\ad[\phi_j]$ seeded by $\phi_0=\chi$, with $\ad[\cdot]\equiv \frac{1}{2}[H_G,\cdot]$, and for the present Hermitian jump operator $\ell=\sqrt{\gamma}\,\chi$, the adjoint Liouvillian acts as
\begin{align}
    \label{eq:Li-adjoint}
    \Li^\dagger[O] = 2\imi \ad[O] + \gamma(\chi O\chi-O).
\end{align}
Writing $\phi_j=r_j\chi+\phi_j^{\perp}$, where $r_j$ is a scalar coefficient and $\{\chi,\phi_j^{\perp}\}=0$, the key identity $\chi\phi_j\chi-\phi_j=-2(\phi_j-r_j\chi)$ gives
\begin{align}
    \Li^\dagger[\phi_j]
    =
    2\imi\phi_{j+1}-2\gamma\qty(\phi_j-r_j\chi).
    \label{eq:Li-phi-j-rj}
\end{align}
Thus the dissipative term maps each Krylov vector back into the span of $\{\phi_0,\phi_1,\dots\}$, so the full adjoint Liouvillian closes on that space.
Introducing the generating functions $\Phi_G(u)\equiv\sum_{j=0}^{\infty}u^j\phi_j$ and
\begin{align}
    R_G(u)
    \equiv
    \sum_{j=0}^{\infty}u^j r_j
    =
    \frac{P_{G \setminus K_s}(u^2)}{P_G(u^2)},
\end{align}
where the second equality follows from Ref.~\cite{fukai-vona-pozsgay-ffd-eigenstates-paper2-2026}.
As shown in Ref.~\cite{fukai-vona-pozsgay-ffd-eigenstates-paper2-2026}, $\Phi_G(u)$ is the unique formal power series satisfying
\begin{align}
    u\ad[\Phi_G(u)]=\Phi_G(u)-\chi
    \label{eq:Krylov-diff-eq}
\end{align}
with $\Phi_G(0)=\chi$.
Summing Eq.~\eqref{eq:Li-phi-j-rj} over $j$ gives
\begin{align}
    \Li^\dagger[\Phi_G(u)]
    =
    2\imi\frac{\Phi_G(u)-\chi}{u}-2\gamma\qty[\Phi_G(u)-R_G(u)\chi].
\end{align}
Rewriting the above as $(\mathbb{1}-z\Li^\dagger)\Phi_G(u) = (1-2\imi z/u+2\gamma z)\Phi_G(u) + (2\imi z/u-2\gamma z R_G(u))\chi$, we see that the choice $u(z)=2\imi z/(1+2\gamma z)$ kills the $\Phi_G$ coefficient, leaving
\begin{align}
    (\mathbb{1}-z\Li^\dagger)\Phi_G(u(z))
    =
    \qty[1+2\gamma z-2\gamma z R_G(u(z))]\chi.
\end{align}
Comparing with the resolvent $\Omega(z)\equiv\sum_{n=0}^{\infty} z^n (\Li^\dagger)^n[\chi]$, which satisfies $(\mathbb{1}-z\Li^\dagger)\Omega(z)=\chi$, we obtain
\begin{align}
    \Omega(z)
    =
    \frac{\Phi_G(u(z))}{1+2\gamma z-2\gamma zR_G(u(z))}.
\end{align}
From $\ev{\phi_j\chi}_{\infty}=r_j$~\cite{fukai-vona-pozsgay-ffd-eigenstates-paper2-2026}, we can see $\ev{\Omega(z)\chi}_{\infty}=R_G(u(z))/[1+2\gamma z-2\gamma z R_G(u(z))]$.
Since $B(t)=\sum_{n=0}^{\infty}\frac{t^n}{n!}\ev{(\Li^\dagger)^n[\chi]\chi}_{\infty}$, term-by-term Laplace transformation gives $\widehat{B}(s)=\sum_{n=0}^{\infty}s^{-(n+1)}\ev{(\Li^\dagger)^n[\chi]\chi}_{\infty}=s^{-1}\ev{\Omega(1/s)\chi}_{\infty}$, and hence
\begin{align}
    \widehat{B}(s)
    &=
    \frac{1}{s}\frac{R_G(u(s^{-1}))}{1+\frac{2\gamma}{s}-\frac{2\gamma}{s}R_G(u(s^{-1}))}
    \nonumber\\
    &=
    \frac{P_{G\setminus K_s}(u(s^{-1})^2)}{(s+2\gamma)P_G(u(s^{-1})^2)-2\gamma P_{G\setminus K_s}(u(s^{-1})^2)}
    \nonumber\\
    &=
    \frac{u(s^{-1})\,P_{G\setminus K_s}(u(s^{-1})^2)}{2\imi\,\polytilpm{G}{+}{u(s^{-1})}}\,.
    \label{eq:Bt-Laplace}
\end{align}
The poles of $\widehat{B}(s)$ occur when $\polytilpm{G}{+}{u(s^{-1})}=0$.
Writing $u(s^{-1})=\liouvillian{u}_k$ and solving for $s$ gives $s=-2\gamma+2\imi/\liouvillian{u}_k$.
Since all poles are simple, the inverse Laplace transform gives
\begin{align}
    B(t)
    &=
    \frac{1}{2\pi\imi}\int_{c-\imi\infty}^{c+\imi\infty}ds\;\widehat{B}(s)\,e^{st}
    \nonumber\\
    &=
    \sum_{k=1}^{\alpha_{\liouvillian{G}}}
    \Res_{s=s_k}\widehat{B}(s)\;e^{s_k t},
    \quad
    s_k\equiv -2\gamma+\frac{2\imi}{\liouvillian{u}_k},
\end{align}
where $c>0$ is chosen so that the Bromwich contour lies to the right of all poles, and the second equality follows by closing the contour to the left.
Using $\sum_k \liouvillian{\varepsilon}_k=-\imi\gamma$ and $\Im \liouvillian{\varepsilon}_k\le0$, we have $-\gamma\le\Im \liouvillian{\varepsilon}_k\le0$ for each $k$, hence $\Re s_k=-2\gamma-2\Im \liouvillian{\varepsilon}_k\le0$.
Evaluating the residue of Eq.~\eqref{eq:Bt-Laplace} at each $s_k$ reproduces Eq.~\eqref{eq:Bt-residue}.
For the boundary-driven Fendley model, the closed-system limit and the uniform thermodynamic-limit analysis are omitted here.

\end{document}